\documentclass[12pt]{article}
\usepackage{longtable,supertabular}
\usepackage{tabularx}
\usepackage{amsmath}
\usepackage{amssymb}
\usepackage{latexsym}
\usepackage{a4wide,color}
\usepackage{multirow}
\usepackage{ntheorem}
\usepackage[square,numbers,sort]{natbib}
\usepackage[title]{appendix}
\usepackage{array}
\usepackage{makecell}

\usepackage[hidelinks]{hyperref}

\newtheorem{theorem}{Theorem}[section]
\newtheorem{corollary}{Corollary}[section]
\newtheorem{proposition}{Proposition}[section]
\newtheorem{lemma}{Lemma}[section]

\newtheorem*{proof}{Proof.}

\newtheorem{example}{Example}[section]

\newtheorem{remark}{Remark}[section]

\title{\bf On True Minimum Distances of Goppa Codes}
\author{Yaqi Chen, Hao Chen, Chao Liu, and Huimin Lao
	 \thanks{Yaqi Chen is with the College of Cyber Security, Jinan University, Guangzhou, Guangdong Province, 510632, China (e-mail: chenyq@stu.jnu.edu.cn).
	 	Hao Chen is with the College of Information Science and Technology, Jinan University, Guangzhou, Guangdong Province, 510632, China (e-mail:
	 	haochen@jnu.edu.cn).
	 	Chao Liu is with the College of Information Science and Technology, Jinan University, Guangzhou, Guangdong Province, 510632, China (chliuu@163.com).
	 	Huimin Lao is with Strategic Centre for Research in Privacy-Preserving Technologies and Systems, Nanyang Technological University, Singapore (e-mail: huimin.lao@ntu.edu.sg).
	 	The research was supported by NSFC Grant 62032009, and the National Research Foundation, Singapore and Infocomm Media Development Authority under its Trust Tech Funding Initiative. Any opinions, findings and conclusions or recommendations expressed in this material are those of the author(s) and do not reflect the views of National Research Foundation, Singapore and Infocomm Media Development Authority.
	 	(Corresponding author: Hao Chen)
	 }
}

\begin{document}
	
	\maketitle
	\begin{abstract}
		Binary Goppa codes are important in coding theory and cryptography. It is well-known that the minimum distance of a binary Goppa code defined by a squarefree polynomial of degree $r$ is at least $2r+1$. Though from the classical results of C. J. Moreno, O. Moreno, F. Levy-dit-Vehel and S. Litsyn that the true minimum distances of binary Goppa codes are $2r+1$, when the extension field degree $m$ is larger than $(r-1)^{4r}$. There have only very few families of binary Goppa codes with determined true minimum distances.\\

In this paper, we give necessary and sufficient conditions for binary Goppa codes such that their minimum distances are $2r+1$ and $2r+2$. Infinitely many infinite families of binary Goppa codes with minimum distance $2r+1$ are given. In particular, we construct such codes for every Goppa degree $r$ in the range $2^{h-1}<r<2^h$ over $\mathbb F_{2^{2h}}$. For every $2\le r\le\lfloor2^m/3\rfloor$, we also obtain at least $2^m-3r+1$ distinct monic squarefree Goppa polynomials of degree $r$ over $\mathbb F_{2^m}$, such that their related binary Goppa codes have minimum distance $2r+1$. We extend a criterion for $q$-ary Goppa codes to split auxiliary polynomials and obtain some infinite families of $q$-ary Goppa codes attaining their designed distances $r+1$. We also construct infinite families of binary Goppa codes with minimum distance exactly $2r+2$.\\

	\end{abstract}
	
	\newpage

\section{Introduction}\label{s1}

\subsection{Background and Motivation}

Goppa codes, introduced by Goppa~\cite{Goppa1970,Goppa1971}, form an important class of alternant codes. Binary Goppa codes can be efficiently decoded~\cite{Patterson1975} and play an important role in code-based cryptography, particularly in the McEliece cryptosystem~\cite{BardetMoraTillich2024}. However, determining their exact minimum distances remains difficult. Apart from the general results of Moreno and Moreno~\cite{MorenoMoreno1992} for sufficiently long codes, exact minimum distances are known mainly for special infinite families. This motivates the study of more general constructions of Goppa codes with known minimum distances.

Throughout this paper, let $q$ be a prime power, and let $\mathbb F_q$ and $\mathbb F_{q^m}$ denote the finite fields with $q$ and $q^m$ elements, respectively. For a vector $\boldsymbol c\in\mathbb F_q^n$, let $\operatorname{supp}(\boldsymbol c)$ and $\operatorname{wt}(\boldsymbol c)$ denote its support and Hamming weight, respectively. A $q$-ary linear code $\mathcal C$ of length $n$, dimension $k$, and minimum distance $d(\mathcal C)$ is denoted by an $[n,k,d]_q$ code, where $d=\min\{\operatorname{wt}(\boldsymbol c):\boldsymbol c\in\mathcal C\setminus\{\boldsymbol0\}\}$.

Let $L=\{\alpha_1,\ldots,\alpha_n\}\subseteq\mathbb F_{q^m}$, and let $G(X)\in\mathbb F_{q^m}[X]$ be a polynomial of degree $r\ge1$ such that $G(\alpha_i)\ne0$ for all $1\le i\le n$. The $q$-ary Goppa code defined by $L$ and $G(X)$ is
\[
\Gamma_q(L,G)=\left\{(c_1,\ldots,c_n)\in\mathbb F_q^n:
\sum_{i=1}^n\frac{c_i}{X-\alpha_i}\equiv0\pmod{G(X)}\right\}.
\]
Here $L$ is the support set and $G(X)$ the Goppa polynomial. The classical Goppa bound gives $d(\Gamma_q(L,G))\ge r+1$. For binary Goppa codes with squarefree $G(X)$, the identity $\Gamma_2(L,G)=\Gamma_2(L,G^2)$ gives the stronger bound
\[
d(\Gamma_2(L,G))\ge2r+1
\]
\cite{SugiyamaKasaharaHirasawaNamekawa1976}.
The bounds $r+1$ and $2r+1$ are the corresponding designed distances. We mainly study binary Goppa codes with minimum distance $2r+1$ or $2r+2$ and also consider constructions of $q$-ary Goppa codes.

\subsection{Related Work}

Early studies of Goppa codes focused on bounds for their minimum distances and special classes of Goppa polynomials~\cite{SugiyamaKasaharaHirasawaNamekawa1976,Feng1983,YueJi1991}.
Moreno and Moreno~\cite{MorenoMoreno1992} used exponential sums to derive a sufficient condition for binary Goppa codes to attain the minimum distance $2r+1$. Let $G(X)\in\mathbb F_{2^m}[X]$ be squarefree of degree $r$, and let $L=\{\alpha\in\mathbb F_{2^m}:G(\alpha)\ne0\}$. Their result, restated in~\cite[Theorem~3]{LevyDitVehelLitsyn1997}, gives $d(\Gamma_2(L,G))=2r+1$ provided
\begin{equation*}
	2^m\ge\max\left\{\left((2r-2)^{2r+1}+1\right)^2,\left(\frac{(2r)!}{r!2^r}\right)^2\right\}.
\end{equation*}
For $r\ge2$, Levy-dit-Vehel and Litsyn~\cite[Theorem~7]{LevyDitVehelLitsyn1997} proved that $d(\Gamma_2(L,G))=2r+1$ provided
\begin{equation*}
	2^m\ge4(\nu+r-1)^{4r+2}\left(\frac{2}{1+\sqrt{1-\frac{1}{(\nu+r-1)^{4r}}}}\right)^{4r+2},
\end{equation*}
where $\nu=2^{-m/2-1}|\{\alpha\in\mathbb F_{2^m}:G(\alpha)=0\}|$. Both results require a sufficiently large field relative to the degree $r$ of the Goppa polynomial $G(X)$. The bound in~\cite[Theorem~7]{LevyDitVehelLitsyn1997} also depends on the number of roots of $G(X)$ in $\mathbb F_{2^m}$.
Finiasz~\cite{Finiasz2003,Finiasz2003extended} studied the search for codewords of weight $2r+1$ in binary Goppa codes. His method uses completely split locator polynomials $F(X)$ of degree $2r+1$ satisfying $F'(X)=G(X)^2$. This provides a way to find minimum-weight codewords, but does not give explicit infinite families or general existence bounds.
Bezzateev and Shekhunova~\cite{BezzateevShekhunova1995} constructed binary Goppa codes with $G(X)=X^t+A$ and minimum distance $2t+1$, where $t\mid 2^m-1$ and $A$ is a $t$-th power in $\mathbb F_{2^m}^*$. They also determined the minimum distances of certain chains of separable binary Goppa codes~\cite{BezzateevShekhunova2008} and improved parameter bounds for totally decomposed cumulative Goppa codes~\cite{BezzateevShekhunova2019}. V\'eron~\cite{Veron1998} studied trace Goppa codes and proved that certain binary Goppa codes have only even-weight codewords. Byrne et al.~\cite{Byrne2023} obtained improved lower bounds on the minimum distances of Goppa codes defined by trace polynomials over extensions of degree at least three.

Chen et al.~\cite{ChenChenDingLao2026b} gave a necessary and sufficient condition for a $q$-ary Goppa code to have minimum distance $r+1$, where $r$ is the degree of its Goppa polynomial. They determined the minimum distances of several families, including certain wild Goppa codes, and extended the binary family $G(X)=X^t+A$ to odd prime powers. Dong et al.~\cite{DongChenChenXieLi2026} studied Goppa polynomials of the form $G(X)=U(X)H(X)+V(X)H'(X)$, where $H(X)$ is irreducible of degree $r+1$. They gave a criterion for the existence of codewords of weight $r+1$ and obtained further families attaining the Goppa bound.
For binary Goppa codes, He et al.~\cite{HeLiQu2027} extended the family $G(X)=X^{3t}+1$ to $G(X)=X^{(2\beta+1)t}+1$ and gave sufficient conditions for the minimum distance to be $(4\beta+2)t+1$. They also constructed three explicit infinite families with $G(X)=X^{3t}+1$ and minimum distance $6t+1$. In another work, He et al.~\cite{HeLiQu2026} gave construction criteria for Goppa polynomials of the forms $G(X)=f(X^t)$ and $G(X)=A(X)h(\phi(X))$. Their results include infinite families with $G(X)=X^{2t}+X^t+1$ and minimum distance $4t+1$, and with $G(X)=X^{3t}+X^t+1$ and minimum distance $6t+1$.

Overall, apart from the result of Moreno and Moreno for sufficiently large fields, exact minimum distances of binary Goppa codes are known mainly for special families. General constructions covering a wide range of Goppa degrees and producing many distinct Goppa polynomials over a fixed finite field remain limited.

\subsection{Main Contributions and Organization}

The main contributions of this paper are summarized as follows.
\begin{enumerate}
	
	\item We give necessary and sufficient conditions for binary Goppa codes to have minimum distance $2r+1$ or $2r+2$ using codeword locator polynomials, as shown in Theorems~\ref{thm:split-locator} and~\ref{thm:even-min-distance}. The criterion for $d=2r+1$ provides a unified characterization of the cases attaining the designed distance in \cite{BezzateevShekhunova1995,BezzateevShekhunova2008,HeLiQu2026,HeLiQu2027}. Proposition~\ref{prop:even-weight} characterizes binary Goppa codes with only even-weight codewords using a trace condition.
	
	\item We develop three general constructions of binary Goppa codes with minimum distance $2r+1$ by deleting roots, taking products of split polynomials, and adjoining two roots to locator polynomials. Theorems~\ref{thm:root-deletion}, \ref{thm:odd-polynomial}, and~\ref{thm:degree-extension} give lower bounds on the number of distinct Goppa polynomials obtained. Corollary~\ref{cor:uniform-existence} guarantees the existence of such polynomials for every degree $2\le r\le\lfloor2^m/3\rfloor$ over $\mathbb F_{2^m}$.

	\item We apply these constructions to obtain numerous explicit infinite families of binary Goppa codes with minimum distance $2r+1$. Propositions~\ref{prop:subspace-deletion}--\ref{prop:two-root-families} give families from subspace polynomials, trace polynomials, and other split polynomials. Corollary~\ref{cor:degree-interval} covers every Goppa degree $2^{h-1}<r<2^h$ over $\mathbb F_{2^{2h}}$.

	\item We extend the construction of Dong et al. to $q$-ary Goppa codes using completely split auxiliary polynomials in Theorem~\ref{thm:split-auxiliary-qary}. Proposition~\ref{prop:dong-norm-family} gives an explicit infinite family with minimum distance $r+1$ for every prime power $q\ge4$.
	
	\item We construct binary Goppa codes with minimum distance exactly $2r+2$ using support sets defined by trace conditions. Proposition~\ref{prop:quartic-even-general} gives a general construction, and Proposition~\ref{cor:four-term-even} provides two explicit infinite families with Goppa degree $r=2^{h-1}+1$.
	
\end{enumerate}

Table~\ref{tab:main-families} lists representative infinite families of Goppa codes with determined minimum distances. The families attaining their designed distances use the full support $L=\{\alpha\in\mathbb F_{q^m}:G(\alpha)\ne0\}$. The family with minimum distance $2r+2$ uses the support defined in \eqref{eq:quartic-even-general-support}.

The remainder of this paper is organized as follows. Section~\ref{sec:preliminaries} recalls some basic results on Goppa codes and polynomials over finite fields. Section~\ref{sec:characterization} gives necessary and sufficient conditions for their minimum distances to equal $2r+1$ or $2r+2$. Section~\ref{sec:constructions} presents three general constructions of binary Goppa codes with minimum distance $2r+1$. Section~\ref{sec:explicit-families} applies these constructions to obtain explicit infinite families. Section~\ref{sec:split-auxiliary} studies $q$-ary Goppa codes using split auxiliary polynomials. Section~\ref{sec:even-distance} constructs binary Goppa codes with minimum distance $2r+2$. Section~\ref{s8} concludes the paper.

\begin{table*}[t]
	\centering
	\caption{Representative infinite families of Goppa codes
		with determined minimum distances constructed in this paper}
	\label{tab:main-families}
	\footnotesize
	\renewcommand{\arraystretch}{1.3}
	\setlength{\tabcolsep}{4pt}
	\begin{tabularx}{\textwidth}{
			|>{\raggedright\arraybackslash}p{0.42\textwidth}
			|>{\raggedright\arraybackslash}X
			|c|l|}
		\hline
		$G(X)$ & Main conditions & $d$ & Reference\\
		\hline
		
		$\displaystyle\sum_{j=0}^{h-2}X^{2^j-1}$
		&
		$h\ge3$, $h\mid m$
		&
		$2^{h-1}-1$
		&
		Prop.~\ref{prop:absolute-trace-family}\\
		\hline
		
		$\displaystyle X^{2^h-1}+bX^{2^{h-1}-1}+1$
		&
		$h\ge2$, $2h\mid m$,
		$b\in\mathbb F_{2^h}^*$
		&
		$2^{h+1}-1$
		&
		Prop.~\ref{prop:relative-trace-family}\\
		\hline
		
		$\displaystyle
		\frac{X^{2^{h-1}+1}+X+a}{X+a^2}$
		&
		$h\ge2$, $3h\mid m$,
		$a^{2^h+1}+a+1=0$
		&
		$2^h+1$
		&
		Prop.~\ref{prop:xw4-families}(i)\\
		\hline
		
		$\displaystyle
		\begin{aligned}
			&X^{2^{h-1}+2}+(1+b)X^{2^{h-1}+1}\\
			&\quad+b^2X^{2^{h-1}}
			+\frac{b}{1+b}X^2+\frac{b^4}{1+b}
		\end{aligned}$
		&
		$h\ge3$, $2h\mid m$,
		$z\in\mathbb F_{2^h}$,
		$b=z^2+z$,
		$b\notin\{0,1\}$
		&
		$2^h+5$
		&
		Prop.~\ref{prop:odd-mixed-family}\\
		\hline
		
		$\displaystyle X^{3v}+cX^{2v}+cX^v+1$
		&
		$v\ge1$ odd,
		$15v\mid2^m-1$,
		$c\in\mathbb F_4\setminus\mathbb F_2$
		&
		$6v+1$
		&
		Prop.~\ref{prop:orbit-explicit}\\
		\hline
		
		$\displaystyle X^{2^{h-1}+1}+X+1$
		&
		$h\ge2$, $3h\mid m$
		&
		$2^h+3$
		&
		Prop.~\ref{prop:two-root-families}(i)\\
		\hline
		
		$\displaystyle X^r+(\rho-1)aX^{r-1}-\rho$
		&
		$q\ge4$, $m\ge2$,
		$r=\dfrac{q^m-1}{q-1}$,
		$a\in\mathbb F_{q^m}^*$, $a^r\ne1$,
		$\rho\in\mathbb F_q\setminus\{0,1\}$
		&
		$r+1$
		&
		Prop.~\ref{prop:dong-norm-family}\\
		\hline
		
		$\displaystyle
		X^r+X^{r-1}+\theta X^{r-2}+(\theta+1)X$
		&
		$r=2^{h-1}+1$, $2h\mid m$,
		$h\equiv3\pmod6$ or
		$h\equiv4\pmod8$
		&
		$2^h+4$
		&
		Prop.~\ref{cor:four-term-even}\\
		\hline
		
	\end{tabularx}
\end{table*}

\section{Preliminaries}\label{sec:preliminaries}

We first recall some basic results on Goppa codes and polynomials over finite fields that will be used in the subsequent constructions.

\subsection{Parity Checks and Dual Codes}\label{subsec:parity-trace}

The Goppa code $\Gamma_q(L,G)$ has the following parity-check matrix over $\mathbb F_{q^m}$~\cite[Ch.~12]{MacWilliamsSloane1977}.

\begin{equation}\label{eq:goppa-parity-check}
	\mathbf H=
	\begin{pmatrix}
		G(\alpha_1)^{-1}&G(\alpha_2)^{-1}&\cdots&G(\alpha_n)^{-1}\\
		\alpha_1G(\alpha_1)^{-1}&\alpha_2G(\alpha_2)^{-1}&\cdots&\alpha_nG(\alpha_n)^{-1}\\
		\vdots&\vdots&&\vdots\\
		\alpha_1^{r-1}G(\alpha_1)^{-1}&\alpha_2^{r-1}G(\alpha_2)^{-1}&\cdots&\alpha_n^{r-1}G(\alpha_n)^{-1}
	\end{pmatrix}.
\end{equation}
A vector $\mathbf c=(c_1,\ldots,c_n)\in\mathbb F_q^n$ belongs to $\Gamma_q(L,G)$ if and only if
\begin{equation}\label{eq:goppa-parity-equations}
	\sum_{i=1}^{n}\frac{c_i\alpha_i^j}{G(\alpha_i)}=0,\qquad 0\le j\le r-1.
\end{equation}
Consequently, the Goppa code $\Gamma_q(L,G)$ has dimension $\dim_{\mathbb F_q}\Gamma_q(L,G)\ge n-mr$.

The trace map from $\mathbb F_{q^m}$ to $\mathbb F_q$ is defined by
$\operatorname{Tr}_{\mathbb F_{q^m}/\mathbb F_q}(\alpha)=\sum_{j=0}^{m-1}\alpha^{q^j}$
for $\alpha\in\mathbb F_{q^m}$.
The trace of a vector is defined coordinatewise. By Delsarte's trace duality~\cite[Thm.~2]{Delsarte1975}, the dual of a binary Goppa code has the following representation.

\begin{lemma}\label{lem:trace-dual}
	Let $\Gamma_2(L,G)$ be a binary Goppa code, where $G(X)\in\mathbb F_{2^m}[X]$ has degree $r\ge1$. Then
	\begin{equation}\label{eq:goppa-trace-dual}
		\Gamma_2(L,G)^\perp=\left\{\left(\operatorname{Tr}_{\mathbb F_{2^m}/\mathbb F_2}\left(\frac{u(\alpha_i)}{G(\alpha_i)}\right)\right)_{i=1}^{n}:u(X)\in\mathbb F_{2^m}[X],\ \deg u(X)<r\right\}.
	\end{equation}
\end{lemma}
\begin{proof}
	The row space of $\mathbf H$ over $\mathbb F_{2^m}$ consists of the vectors $(u(\alpha_i)/G(\alpha_i))_{i=1}^{n}$, where $u(X)\in\mathbb F_{2^m}[X]$ and $\deg u(X)<r$. Since $\Gamma_2(L,G)$ is the subfield subcode of the kernel of $\mathbf H$ over $\mathbb F_{2^m}$, Delsarte's trace duality gives \eqref{eq:goppa-trace-dual}.
\end{proof}

Let $\mathbf1=(1,\ldots,1)\in\mathbb F_2^n$. Since $\mathbf c\cdot\mathbf1=\operatorname{wt}(\mathbf c)\pmod2$, every codeword of $\Gamma_2(L,G)$ has even weight if and only if $\mathbf1\in\Gamma_2(L,G)^\perp$.

\subsection{Locator Polynomials and Interpolation}\label{subsec:polynomial-tools}

Let $S\subseteq L$ be nonempty, and define $F(X)=\prod_{\alpha\in S}(X-\alpha)$. Then $F(X)$ is monic and squarefree, with $\deg F(X)=|S|$, and
\begin{equation}\label{eq:locator-logarithmic-derivative}
	\frac{F'(X)}{F(X)}=\sum_{\alpha\in S}\frac1{X-\alpha}.
\end{equation}
Over $\mathbb F_{2^m}$, the derivative of every polynomial is a square, and a polynomial has zero derivative if and only if it is a square.

\begin{lemma}[Lagrange interpolation]\label{lem:lagrange}
	Let $S=\{\beta_1,\ldots,\beta_t\}\subseteq L$ be nonempty, and let $F(X)=\prod_{i=1}^{t}(X-\beta_i)$. For every $R(X)\in\mathbb F_{q^m}[X]$ with $\deg R(X)<t$,
	\begin{equation}\label{eq:lagrange-interpolation}
		R(X)=\sum_{i=1}^{t}R(\beta_i)\frac{F(X)}{(X-\beta_i)F'(\beta_i)}.
	\end{equation}
	If $t\ge2$, then
	\begin{equation}\label{eq:lagrange-moment-identity}
		\sum_{i=1}^{t}\frac{\beta_i^j}{F'(\beta_i)}=0,\qquad 0\le j\le t-2.
	\end{equation}
\end{lemma}

\begin{proof}
	The polynomial $F(X)/((X-\beta_i)F'(\beta_i))$ has degree $t-1$ and takes the value $1$ at $\beta_i$ and $0$ at every $\beta_j$ with $j\ne i$. The right-hand side of \eqref{eq:lagrange-interpolation} has degree less than $t$ and agrees with $R(X)$ at all $t$ elements of $S$. This proves \eqref{eq:lagrange-interpolation}. Setting $R(X)=X^j$ and comparing the coefficients of $X^{t-1}$ gives \eqref{eq:lagrange-moment-identity}.
\end{proof}

\subsection{Polynomials over Finite Fields}\label{subsec:splitting}

\begin{lemma}\label{lem:finite-field-splitting}
	Let $h\mid m$, and let $P(X)\in\mathbb F_{q^h}[X]$ be nonconstant. Then $P(X)$ splits into distinct linear factors over $\mathbb F_{q^m}$ if and only if $P(X)\mid X^{q^m}-X$. Equivalently, $P(X)$ is squarefree and every irreducible factor of degree $d$ over $\mathbb F_{q^h}$ satisfies $hd\mid m$.
\end{lemma}

\begin{proof}
	The first equivalence follows from $X^{q^m}-X=\prod_{\alpha\in\mathbb F_{q^m}}(X-\alpha)$. The second follows since an irreducible polynomial of degree $d$ over $\mathbb F_{q^h}$ splits over $\mathbb F_{q^m}$ if and only if $hd\mid m$.
\end{proof}

The polynomial $X^s-1$ splits into distinct linear factors over $\mathbb F_{q^m}$ if and only if $s\mid q^m-1$.

\begin{lemma}\label{lem:subspace-polynomial}
	Let $1\le h\le m$, and let $V$ be an $h$-dimensional $\mathbb F_2$-subspace of $\mathbb F_{2^m}$. The subspace polynomial of $V$ is a monic linearized polynomial of the form
	\[
	P(X)=\prod_{v\in V}(X-v)=\sum_{j=0}^{h}c_jX^{2^j},
	\]
	where $c_j\in\mathbb F_{2^m}$, $c_h=1$, and $c_0\ne0$. In particular, $P'(X)=c_0$.
\end{lemma}

We also use the transitivity of the trace map. For $h\mid m$ and $\alpha\in\mathbb F_{2^m}$,
\[
\operatorname{Tr}_{\mathbb F_{2^m}/\mathbb F_2}(\alpha)
=
\operatorname{Tr}_{\mathbb F_{2^h}/\mathbb F_2}
\left(\operatorname{Tr}_{\mathbb F_{2^m}/\mathbb F_{2^h}}(\alpha)\right).
\]

\begin{lemma}\label{lem:quadratic-splitting}
	Let $h\ge1$, and let $P(X)=X^2+bX+c\in\mathbb F_{2^h}[X]$, where $b\ne0$. Then $P(X)$ splits over $\mathbb F_{2^h}$ if and only if
	\[
	\operatorname{Tr}_{\mathbb F_{2^h}/\mathbb F_2}\left(\frac{c}{b^2}\right)=0.
	\]
	If this trace equals $1$, then $P(X)$ is irreducible over $\mathbb F_{2^h}$, and its roots belong to $\mathbb F_{2^{2h}}$. Each root $\alpha$ satisfies
	\[
	\operatorname{Tr}_{\mathbb F_{2^{2h}}/\mathbb F_{2^h}}(\alpha)=\alpha^{2^h}+\alpha=b.
	\]
\end{lemma}

\begin{proof}
	Substituting $X=bZ$ into $P(X)=0$ gives $Z^2+Z=c/b^2$. The $\mathbb F_2$-linear map $z\mapsto z^2+z$ on $\mathbb F_{2^h}$ has kernel $\{0,1\}$ and image of cardinality $2^{h-1}$. Since $\operatorname{Tr}_{\mathbb F_{2^h}/\mathbb F_2}(z^2+z)=0$, its image equals the kernel of the absolute trace map. This proves the splitting criterion. The roots are distinct since $P'(X)=b\ne0$. If $P(X)$ is irreducible, its roots in $\mathbb F_{2^{2h}}$ are $\alpha$ and $\alpha^{2^h}$, giving $\alpha+\alpha^{2^h}=b$.
\end{proof}

\section{Codeword Structure and Minimum Distances of Binary Goppa Codes}\label{sec:characterization}

Throughout this section, let $m\ge1$, and let $G(X)\in\mathbb F_{2^m}[X]$ be monic and squarefree of degree $r\ge1$. Let $L\subseteq\{\alpha\in\mathbb F_{2^m}:G(\alpha)\ne0\}$. We assume that $\Gamma_2(L,G)$ is nonzero whenever its minimum distance is considered.
We characterize binary Goppa codewords by locator polynomials and give necessary and sufficient conditions for $\Gamma_2(L,G)$ to have minimum distance $2r+1$ or $2r+2$.

\begin{proposition}\label{prop:codeword-derivative}
	Let $S\subseteq L$ be nonempty, and define $F(X)=\prod_{\alpha\in S}(X-\alpha)$. Then $S$ is the support of a codeword in $\Gamma_2(L,G)$ if and only if $G(X)^2\mid F'(X)$. The codeword has weight $|S|=\deg F(X)$.
\end{proposition}

\begin{proof}
	By \eqref{eq:locator-logarithmic-derivative}, the binary vector with support $S$ belongs to $\Gamma_2(L,G)$ if and only if
	\[
	\frac{F'(X)}{F(X)}\equiv0\pmod{G(X)}.
	\]
	Since $\gcd(F(X),G(X))=1$, this is equivalent to $G(X)\mid F'(X)$. As $F'(X)$ is a square polynomial over $\mathbb F_{2^m}$ and $G(X)$ is squarefree, $G(X)\mid F'(X)$ if and only if $G(X)^2\mid F'(X)$.
\end{proof}

Since $F(X)$ is squarefree, $F'(X)\ne0$.
Proposition~\ref{prop:codeword-derivative} gives $2r\le\deg F'(X)\le\deg F(X)-1$ for every nonzero codeword.
Hence $d(\Gamma_2(L,G))\ge2r+1$.
The proposition also gives a bijection between nonzero codewords and monic divisors $F(X)$ of $\prod_{\alpha\in L}(X-\alpha)$ satisfying $G(X)^2\mid F'(X)$.

\begin{theorem}\label{thm:split-locator}
	The minimum distance of $\Gamma_2(L,G)$ is $2r+1$ if and only if there exist pairwise distinct elements $\alpha_{i_1},\ldots,\alpha_{i_{2r+1}}\in L$ such that
	\begin{equation}\label{eq:binary-ratio-criterion}
		\frac{G(\alpha_{i_j})^2}{G(\alpha_{i_{2r+1}})^2}
		\frac{F'(\alpha_{i_{2r+1}})}{F'(\alpha_{i_j})}=1,
		\qquad 1\le j\le2r,
	\end{equation}
	where $F(X)=\prod_{\ell=1}^{2r+1}(X-\alpha_{i_\ell})$.
	Equivalently, there exists a monic polynomial $F(X)\in\mathbb F_{2^m}[X]$ of degree $2r+1$ with distinct roots in $L$ such that $F'(X)=G(X)^2$.
\end{theorem}

\begin{proof}
	Suppose that $d(\Gamma_2(L,G))=2r+1$, and let $F(X)$ be the monic locator polynomial of a codeword of weight $2r+1$. By Proposition~\ref{prop:codeword-derivative},
	$G(X)^2\mid F'(X)$. Since $F(X)$ is monic and squarefree of degree $2r+1$, its derivative is monic of degree $2r$.
	Hence $F'(X)=G(X)^2$, which implies \eqref{eq:binary-ratio-criterion}.
	
	Conversely, suppose that \eqref{eq:binary-ratio-criterion}
	holds. Let
	$\lambda=G(\alpha_{i_{2r+1}})^2/F'(\alpha_{i_{2r+1}})$.
	Then $G(\alpha_{i_j})^2=\lambda F'(\alpha_{i_j})$ for all $1\le j\le2r+1$.
	Since $G(X)^2-\lambda F'(X)$ has degree at most $2r$ and vanishes at $2r+1$ distinct points,
	$G(X)^2=\lambda F'(X)$.
	Comparing leading coefficients gives $\lambda=1$. Proposition~\ref{prop:codeword-derivative} now yields $d(\Gamma_2(L,G))=2r+1$.
\end{proof}

\begin{corollary}\label{cor:split-locator}
	The minimum distance of $\Gamma_2(L,G)$ is $2r+1$ if and only if there exists $u(X)\in\mathbb F_{2^m}[X]$ with $\deg u(X)\le r$ such that
	\begin{equation}\label{eq:odd-normal-form-criterion}
		XG(X)^2+u(X)^2\mid\prod_{\alpha\in L}(X-\alpha).
	\end{equation}
	For the full support $L=\{\alpha\in\mathbb F_{2^m}:G(\alpha)\ne0\}$, this condition is equivalent to $XG(X)^2+u(X)^2\mid X^{2^m}-X$.
\end{corollary}

\begin{proof}
	Every polynomial $F(X)$ satisfying $F'(X)=G(X)^2$
	can be written uniquely as $F(X)=XG(X)^2+u(X)^2$. Since $G(X)$ is monic of degree $r$, the polynomial $F(X)$ is monic of degree $2r+1$ if and only if $\deg u(X)\le r$.
	
	For the full support, Lemma~\ref{lem:finite-field-splitting} implies that $F(X)\mid X^{2^m}-X$ if and only if $F(X)$ has distinct roots in $\mathbb F_{2^m}$. Since $F'(X)=G(X)^2$, the squarefreeness of $F(X)$ implies $\gcd(F(X),G(X))=1$. All roots of $F(X)$ therefore belong to $L$. Conversely, $F(X)\mid\prod_{\alpha\in L}(X-\alpha)$ implies $F(X)\mid X^{2^m}-X$.
\end{proof}

Theorem~\ref{thm:split-locator} also provides a method for constructing binary Goppa codes attaining the designed distance. Let $F(X)\in\mathbb F_{2^m}[X]$ be monic of degree $2r+1$ and split into distinct linear factors over $\mathbb F_{2^m}$. If $F'(X)=G(X)^2$ for a squarefree polynomial $G(X)$, then $d(\Gamma_2(L,G))=2r+1$ for every support $L$ containing all roots of $F(X)$.

\begin{remark}\label{R-HLQ}
	
	The constructions in \cite[Theorems~3.2 and~3.5]{HeLiQu2026} and \cite[Theorem~3.1]{HeLiQu2027} satisfy the locator criterion in Theorem~\ref{thm:split-locator}. In each construction, the selected $2r+1$ support elements define a monic locator polynomial $F(X)$ satisfying $G(X)\mid F'(X)$. Since $G(X)$ is squarefree, Proposition~\ref{prop:codeword-derivative} gives $G(X)^2\mid F'(X)$. Comparing degrees and leading coefficients gives $F'(X)=G(X)^2$.

	In \cite[Theorem~3.2]{HeLiQu2026}, $G(X)=f(X^t)$, where $t$ is odd and $f(X)$ is monic and squarefree of degree $s$. The monic auxiliary polynomial $P(X)$ has degree $2s$ and satisfies $f(X)\mid P(X)+XP'(X)$. Since $P(X)+XP'(X)$ is a square polynomial, it equals $f(X)^2$. Thus $F(X)=XP(X^t)$ satisfies $F'(X)=P(X^t)+X^tP'(X^t)=G(X)^2$. This includes $G(X)=X^{2t}+X^t+1$ and $G(X)=X^{3t}+X^t+1$ in \cite[Propositions~3.3 and~3.4]{HeLiQu2026}.
	
	In \cite[Theorem~3.5]{HeLiQu2026}, $G(X)$ is the monic normalization of $A(X)h(\phi(X))$. Writing $Q(Y)=\prod_{j=1}^{N}(Y-u_j)$ and denoting the leading coefficient of $\phi(X)$ by $a$, the monic locator is $F(X)=a^{-N}Q(\phi(X))$. The conditions $h(Y)\mid Q'(Y)$ and $A(X)\mid\phi'(X)$ give $G(X)\mid F'(X)$. This includes $G(X)=X^4+cX$ in \cite[Theorem~3.8]{HeLiQu2026}.
	
	In \cite[Theorem~3.1]{HeLiQu2027}, $G(X)=X^{(2\beta+1)t}+1$. For the elements $x_1,\ldots,x_{2\beta+1}$ specified there, the locator is $F(X)=XQ(X^t)$, where $Q(Y)=\prod_{j=1}^{2\beta+1}(Y-x_j^t)(Y-x_j^{-t})$. The equations satisfied by $x_j$ imply $G(X)\mid F'(X)$. Comparing degrees and leading coefficients gives $F'(X)=G(X)^2$. This includes the families with $G(X)=X^{3t}+1$ in \cite[Corollaries~4.2, 4.4, and~4.6]{HeLiQu2027}.
	
\end{remark}

\begin{remark}\label{rem:good-zero-sets}
	Primitive narrow-sense BCH codes are special cases of Goppa codes with $G(X)=X^t$ and $L=\mathbb F_{q^m}^*$~\cite[Ch.~12, Sec.~3]{MacWilliamsSloane1977}. Zheng's $\mathbb F_2$-good zero-sets contain $0$, which is excluded from the BCH codeword support~\cite{Zheng2026}. For binary Goppa codes, the root set $S$ of a locator polynomial $F(X)$ is itself the codeword support and may contain $0$. When $|S|=2r+1$, the condition $F'(X)=G(X)^2$ gives a codeword of weight $2r+1$, provided $G(X)=\sqrt{F'(X)}$ is squarefree. If $0\notin S$ and $F(X)=XG(X)^2+u(X)^2$, then $S\cup\{0\}$ is an $\mathbb F_2$-good zero-set if and only if $u(X)$ is a nonzero constant. Thus the two criteria are related but not equivalent.
\end{remark}

We next characterize the binary Goppa codes with the minimum distance $2r+2$.

\begin{theorem}\label{thm:even-min-distance}
	The minimum distance of $\Gamma_2(L,G)$ is $2r+2$ if and only if both of the following conditions hold:
	\begin{enumerate}
		\item[(i)] There is no $u(X)\in\mathbb F_{2^m}[X]$ with $\deg u(X)\le r$ such that $XG(X)^2+u(X)^2\mid\prod_{\alpha\in L}(X-\alpha)$.
		\item[(ii)] There exist a monic divisor $F(X)$ of $\prod_{\alpha\in L}(X-\alpha)$ of degree $2r+2$ and $\lambda\in\mathbb F_{2^m}^*$ such that $F'(X)=\lambda G(X)^2$.
	\end{enumerate}
	Condition (ii) is equivalent to the existence of $\lambda\in\mathbb F_{2^m}^*$ and a monic polynomial $u(X)\in\mathbb F_{2^m}[X]$ of degree $r+1$ such that
	\begin{equation}\label{eq:even-normal-form-criterion}
		\lambda XG(X)^2+u(X)^2\mid\prod_{\alpha\in L}(X-\alpha).
	\end{equation}
	If every codeword of $\Gamma_2(L,G)$ has even weight, condition (ii) alone is equivalent to $d(\Gamma_2(L,G))=2r+2$.
\end{theorem}

\begin{proof}
	By Corollary~\ref{cor:split-locator}, condition (i) is equivalent to the absence of codewords of weight $2r+1$.
	
	Let $F(X)$ be a monic divisor of $\prod_{\alpha\in L}(X-\alpha)$ of degree $2r+2$. By Proposition~\ref{prop:codeword-derivative}, $F(X)$ is the locator polynomial of a codeword if and only if $G(X)^2\mid F'(X)$. Since $F(X)$ is squarefree of even degree, $F'(X)\ne0$ and $\deg F'(X)\le2r$. Consequently, $G(X)^2\mid F'(X)$ if and only if $F'(X)=\lambda G(X)^2$ for some $\lambda\in\mathbb F_{2^m}^*$. Since $d(\Gamma_2(L,G))\ge2r+1$, conditions (i) and (ii) together are equivalent to $d(\Gamma_2(L,G))=2r+2$.
	
	The identity $F'(X)=\lambda G(X)^2$ holds if and only if $F(X)=\lambda XG(X)^2+u(X)^2$ for a unique polynomial $u(X)\in\mathbb F_{2^m}[X]$. Since $\deg F(X)=2r+2$ and $F(X)$ is monic, $u(X)$ is monic of degree $r+1$. If every codeword of $\Gamma_2(L,G)$ has even weight, condition (i) holds automatically.
\end{proof}

The trace description in Lemma~\ref{lem:trace-dual} characterizes binary Goppa codes whose codewords all have even weight.

\begin{proposition}\label{prop:even-weight}
	Every codeword of $\Gamma_2(L,G)$ has even weight if and only if there exists $u(X)\in\mathbb F_{2^m}[X]$ with $\deg u(X)<r$ such that
	\begin{equation}\label{eq:even-weight-trace-criterion}
		\operatorname{Tr}_{\mathbb F_{2^m}/\mathbb F_2}\left(\frac{u(\alpha)}{G(\alpha)}\right)=1,\qquad \alpha\in L.
	\end{equation}
\end{proposition}

\begin{proof}
	Every codeword of $\Gamma_2(L,G)$ has even weight if and only if $\mathbf1\in\Gamma_2(L,G)^\perp$. By Lemma~\ref{lem:trace-dual}, this is equivalent to \eqref{eq:even-weight-trace-criterion}.
\end{proof}

\section{Constructions of Binary Goppa Codes with Minimum Distance $2r+1$}\label{sec:constructions}

By Corollary~\ref{cor:split-locator}, the construction of binary Goppa codes with minimum distance $2r+1$ reduces to finding a monic polynomial $F(X)\in\mathbb F_{2^m}[X]$ of degree $2r+1$ such that $F(X)\mid X^{2^m}-X$ and $F'(X)=G(X)^2$, where $G(X)$ is monic and squarefree. We consider root deletion, products of split polynomials, and multiplication by two linear factors. For each Goppa polynomial $G(X)$, let $L=\{\alpha\in\mathbb F_{2^m}:G(\alpha)\ne0\}$.

\subsection{Constructions from Root Deletion}\label{sec:deletion}

We first construct Goppa polynomials by deleting roots from completely split polynomials with nonzero constant derivative.

\begin{theorem}\label{thm:root-deletion}
	Let $m\ge1$, and let $B(X)\in\mathbb F_{2^m}[X]$ be monic of degree $r+1$, where $r\ge1$. Let $\lambda\in\mathbb F_{2^m}^*$. Suppose that $B'(X)\ne0$ and that
	\[
	H(X)=\lambda X+B(X)^2
	\]
	splits over $\mathbb F_{2^m}$. For every root $\alpha$ of $H(X)$, define
	\[
	G(X)=\frac{B(X)-B(\alpha)}{X-\alpha}.
	\]
	Then $G(X)$ is monic of degree $r$. For at least $2r+2-\lfloor r/2\rfloor$ roots $\alpha$, the polynomial $G(X)$ is squarefree and $d(\Gamma_2(L,G))=2r+1$. Distinct roots give distinct Goppa polynomials.
\end{theorem}

\begin{proof}
	Since $H'(X)=\lambda\ne0$, the polynomial $H(X)$ has $2r+2$ distinct roots in $\mathbb F_{2^m}$. Fix a root $\alpha$. Then $B(\alpha)^2=\lambda\alpha$, and
	\[
	F(X)=\frac{H(X)}{X-\alpha}
	=\lambda+(X-\alpha)G(X)^2.
	\]
	Thus $F(X)$ is monic of degree $2r+1$, splits into distinct linear factors over $\mathbb F_{2^m}$, and satisfies $F'(X)=G(X)^2$. Since $\lambda\ne0$, we have $\gcd(F(X),G(X))=1$, so all roots of $F(X)$ belong to $L$. If $G(X)$ is squarefree, Theorem~\ref{thm:split-locator} gives $d(\Gamma_2(L,G))=2r+1$.
	
	Suppose that $G(X)$ has a multiple root $\beta$ in an algebraic closure of $\mathbb F_{2^m}$. From $B(X)-B(\alpha)=(X-\alpha)G(X)$, we obtain $B(\beta)=B(\alpha)$ and $B'(\beta)=0$. Since $H(\alpha)=0$, we have
	\[
	\alpha=\lambda^{-1}B(\beta)^2.
	\]
	Hence each root $\beta$ of $B'(X)$ corresponds to at most one root $\alpha$ for which $G(X)$ is not squarefree. As $B'(X)$ is a nonzero square polynomial of degree at most $r$, it has at most $\lfloor r/2\rfloor$ distinct roots. Therefore, at least $2r+2-\lfloor r/2\rfloor$ deletions give squarefree Goppa polynomials.
	
	It remains to show that these polynomials are distinct. Suppose that two roots $\alpha\ne\gamma$ give the same polynomial $G(X)$. Then
	\[
	B(X)-B(\alpha)=(X-\alpha)G(X),\qquad
	B(X)-B(\gamma)=(X-\gamma)G(X),
	\]
	which implies $(\alpha-\gamma)G(X)=B(\alpha)-B(\gamma)$. This contradicts $\deg G(X)=r\ge1$.
\end{proof}

\begin{example}\label{ex:root-deletion}
	We give two examples of Theorem~\ref{thm:root-deletion} with odd and even Goppa degrees.
	\begin{enumerate}
		\item[(i)] Let $m=5$, $B(X)=X^6+X^5+X^3+X^2+X$, and $\lambda=1$. Then $H(X)=X+B(X)^2$ splits completely over $\mathbb F_{32}$. Deleting the root $a=0$ gives $G(X)=X^5+X^4+X^2+X+1$. The resulting Goppa code has parameters $[27,2,11]_2$.
		\item[(ii)] Let $m=8$, $B(X)=X^7+X^4+X+1$, and $\lambda=1$. Choose $\omega\in\mathbb F_4$ satisfying $\omega^2+\omega+1=0$. The polynomial $H(X)=X+B(X)^2$ splits completely over $\mathbb F_{256}$. Deleting the root $a=\omega$ gives $G(X)=X^6+\omega X^5+\omega^2X^4+1$. The resulting Goppa code has parameters $[255,207,13]_2$.
	\end{enumerate}
\end{example}

If $B'(X)$ is a nonzero constant, every root deletion is successful.
We next consider $B(X)=XW(X)^2$, for which $B'(X)=W(X)^2$ and $H(X)=X(XW(X)^4+1)$.

\begin{proposition}\label{prop:xw4-deletion}
	Let $m\ge1$, and let $W(X)\in\mathbb F_{2^m}[X]$ be monic of degree $t\ge1$. Suppose that
	\[
	P(X)=XW(X)^4+1
	\]
	splits over $\mathbb F_{2^m}$. For every $a\in\mathbb F_{2^m}$ satisfying $P(a^2)=0$, the polynomial
	\[
	G(X)=\frac{XW(X)^2+a}{X-a^2}
	\]
	is monic and squarefree of degree $2t$, and $d(\Gamma_2(L,G))=4t+1$. This construction gives exactly $4t+1$ distinct Goppa polynomials.
\end{proposition}

\begin{proof}
	Let $B(X)=XW(X)^2$ and $H(X)=XP(X)=X+B(X)^2$. Since $H'(X)=1$, the polynomial $H(X)$ has $4t+2$ distinct roots in $\mathbb F_{2^m}$. Deleting the root $\alpha=0$ gives $B(X)/X=W(X)^2$, which is not squarefree.
	
	Let $a\in\mathbb F_{2^m}$ satisfy $P(a^2)=0$. Since $P(0)=1$, we have $a\ne0$. The equality $H(a^2)=0$ gives $B(a^2)^2=a^2$, so $B(a^2)=a$. Hence
	\[
	G(X)=\frac{B(X)-B(a^2)}{X-a^2}.
	\]
	Since $B(a^2)=a$, we have $G(X)=(XW(X)^2+a)/(X-a^2)$. The numerator has derivative $W(X)^2$. Since $a\ne0$, we have $\gcd(XW(X)^2+a,W(X)^2)=1$. Thus the numerator is squarefree, and so is $G(X)$.
	
	Moreover,
	\[
	F(X)=\frac{H(X)}{X-a^2}
	=1+(X-a^2)G(X)^2
	\]
	has $4t+1$ distinct roots in $\mathbb F_{2^m}$ and satisfies $F'(X)=G(X)^2$. Since $\gcd(F(X),G(X))=1$, all roots of $F(X)$ belong to $L$. Theorem~\ref{thm:split-locator} gives $d(\Gamma_2(L,G))=4t+1$.
	
	Finally, $P(X)$ has $4t+1$ distinct roots, and the map $a\mapsto a^2$ is a bijection on $\mathbb F_{2^m}$. Thus there are exactly $4t+1$ admissible values of $a$. By Theorem~\ref{thm:root-deletion}, they give distinct Goppa polynomials.
\end{proof}

\begin{example}\label{ex:xw4-basic}
	Let $m=6$ and $W(X)=X^2$. Since $9\mid2^6-1$, the polynomial $P(X)=X^9+1$ splits over $\mathbb F_{64}$. Taking $a=1$ in Proposition~\ref{prop:xw4-deletion} gives
	\[
	G(X)=\frac{X^5+1}{X+1}
	=X^4+X^3+X^2+X+1.
	\]
	The resulting Goppa code has parameters $[64,40,9]_2$.
\end{example}

\subsection{Constructions from Products of Split Polynomials}

We next construct locator polynomials as products of split polynomials with disjoint root sets.
Consider $F(X)=(X^{2^h}+X+c^2)P(X)$, where $P(X)$ splits over $\mathbb F_{2^h}$.

\begin{theorem}\label{thm:odd-polynomial}
	Let $m\ge1$ and $h\ge2$ with $2h\mid m$, and let $1\le s<2^{h-1}$. Let $P(X)\in\mathbb F_{2^h}[X]$ be monic and squarefree of degree $2s+1$ and split over $\mathbb F_{2^h}$. Write
	\[
	P(X)=A(X)^2+XB(X)^2,
	\]
	where $A(X),B(X)\in\mathbb F_{2^h}[X]$, and suppose that $A'(X)$ and $B'(X)$ are not both zero. Then at least $2^h-s$ values of $c\in\mathbb F_{2^h}^*$ give a monic squarefree polynomial
	\[
	G(X)=A(X)+(X^{2^{h-1}}+c)B(X)
	\]
	of degree $r=2^{h-1}+s$ such that $d(\Gamma_2(L,G))=2r+1=2^h+2s+1$. Distinct such values of $c$ give distinct Goppa polynomials.
\end{theorem}

\begin{proof}
	Since $P(X)$ is monic of degree $2s+1$, the polynomial $B(X)$ is monic of degree $s$, and $\deg A(X)\le s$. Moreover, $P'(X)=B(X)^2$ and $P(X)$ is squarefree, so $\gcd(A(X),B(X))=1$.
	
	For $c\in\mathbb F_{2^h}^*$, let
	\[
	F(X)=(X^{2^h}+X+c^2)P(X).
	\]
	The relative trace map $x\mapsto x^{2^h}+x$ from $\mathbb F_{2^{2h}}$ onto $\mathbb F_{2^h}$ has kernel $\mathbb F_{2^h}$. Hence $X^{2^h}+X+c^2$ has $2^h$ distinct roots in $\mathbb F_{2^{2h}}$, none of which lies in $\mathbb F_{2^h}$. Since $P(X)$ is squarefree and splits over $\mathbb F_{2^h}$, the polynomial $F(X)$ is monic and squarefree of degree $2^h+2s+1$ and splits over $\mathbb F_{2^m}$. We have
	\[
	\begin{aligned}
		F'(X)&=P(X)+(X^{2^h}+X+c^2)B(X)^2\\
		&=A(X)^2+(X^{2^h}+c^2)B(X)^2\\
		&=\bigl(A(X)+(X^{2^{h-1}}+c)B(X)\bigr)^2\\
		&=G(X)^2.
	\end{aligned}
	\]
	Thus $G(X)$ is monic of degree $r=2^{h-1}+s$.
	
	We next determine the values of $c$ for which $G(X)$ is squarefree. Let
	\[
	D(X)=A'(X)B(X)+A(X)B'(X).
	\]
	If $D(X)=0$, then $\gcd(A(X),B(X))=1$ implies $B(X)\mid B'(X)$. Hence $B'(X)=0$, and consequently $A'(X)=0$, contrary to the assumption. Thus $D(X)\ne0$. Since $D'(X)=0$ and $\deg D(X)\le2s-1$, the polynomial $D(X)$ is a nonzero square and has at most $s-1$ distinct roots.
	
	Suppose that $G(X)$ has a multiple root $\beta$ in an algebraic closure of $\mathbb F_{2^m}$. Since $\gcd(A(X),B(X))=1$, we have $B(\beta)\ne0$. The equations $G(\beta)=G'(\beta)=0$ give
	\[
	D(\beta)=0,\qquad
	c=\frac{A(\beta)}{B(\beta)}+\beta^{2^{h-1}}.
	\]
	Thus each root of $D(X)$ determines at most one value of $c$ for which $G(X)$ has a multiple root. At most $s-1$ of the $2^h-1$ nonzero values of $c$ are unsuccessful. Therefore, at least $2^h-s$ values give squarefree Goppa polynomials.
	
	For each such value, $F(X)$ is squarefree and $F'(X)=G(X)^2$, so $F(X)$ and $G(X)$ are coprime. Every root of $F(X)$ therefore belongs to $L$. Theorem~\ref{thm:split-locator} gives $d(\Gamma_2(L,G))=2^h+2s+1$. Since $B(X)\ne0$, distinct values of $c$ give distinct Goppa polynomials.
\end{proof}

The assumption on $A'(X)$ and $B'(X)$ holds automatically when $s$ is odd, since $B'(X)\ne0$. It cannot be omitted in general. For example, over $\mathbb F_{16}$, take $P(X)=X^5+1=1+X(X^2)^2$. Then $G(X)=X^{10}+cX^2+1$ satisfies $G'(X)=0$ for every $c\in\mathbb F_{16}^*$.

\begin{corollary}\label{cor:degree-interval}
	Let $m\ge1$ and $h\ge2$ with $2h\mid m$. For every integer $r$ satisfying $2^{h-1}<r<2^h$, there exist at least $3\cdot2^{h-1}-r$ distinct monic squarefree polynomials $G(X)\in\mathbb F_{2^h}[X]$ of degree $r$ such that $d(\Gamma_2(L,G))=2r+1$.
\end{corollary}

\begin{proof}
	Let $s=r-2^{h-1}$, so $1\le s<2^{h-1}$. Choose $S\subseteq\mathbb F_{2^h}$ with $|S|=2s+1$, and write
	\[
	P(X)=\prod_{\alpha\in S}(X-\alpha)=A(X)^2+XB(X)^2.
	\]
	Then $P(X)$ is monic and squarefree and splits over $\mathbb F_{2^h}$. The polynomial $B(X)$ is monic of degree $s$, and $\gcd(A(X),B(X))=1$.
	
	We show that $S$ can be chosen so that $B'(X)\ne0$. If $s$ is odd, this follows from $\deg B(X)=s$. Suppose that $s$ is even. We may choose $S$ so that the coefficient of $X^{2s-1}$ in $P(X)$ is nonzero. Indeed, if this coefficient is zero, at most one $\alpha\in S$ satisfies
	\[
	\sum_{\gamma\in S\setminus\{\alpha\}}\gamma=0.
	\]
	Choose $\alpha\in S$ for which this sum is nonzero. Since $2s+1<2^h$, there exists $\beta\in\mathbb F_{2^h}\setminus S$. Replacing $\alpha$ by $\beta$ changes the coefficient of $X^{2s-1}$ by
	\[
	(\alpha+\beta)\sum_{\gamma\in S\setminus\{\alpha\}}\gamma\ne0.
	\]
	For the resulting polynomial $P(X)$, the coefficient of $X^{s-1}$ in $B(X)$ is nonzero. Since $s-1$ is odd, we have $B'(X)\ne0$.
	
	By Theorem~\ref{thm:odd-polynomial}, the constructed $P(X)$ gives at least $2^h-s=3\cdot2^{h-1}-r$ distinct monic squarefree Goppa polynomials of degree $r$, each defining a binary Goppa code with minimum distance $2r+1$.
\end{proof}

For every integer $r\ge3$ that is not a power of two, taking $h=\lceil\log_2 r\rceil$ gives $2^{h-1}<r<2^h$. Corollary~\ref{cor:degree-interval} therefore applies whenever $2h\mid m$.

\subsection{Constructions by Adjoining Two Roots}\label{sec:extension}

We finally construct new locator polynomials by adjoining two roots. Let $F(X)$ have $2r+1$ distinct roots in $\mathbb F_{2^m}$, and consider
\[
\widetilde F(X)=(X+\xi^2)(X+\eta^2)F(X),
\]
where $\xi^2$ and $\eta^2$ are distinct and lie outside the root set of $F(X)$.

\begin{theorem}\label{thm:degree-extension}
	Let $m\ge1$ and $r\ge1$, and let $F(X)\in\mathbb F_{2^m}[X]$ be monic of degree $2r+1$ such that $F(X)\mid X^{2^m}-X$. Write
	\[
	F(X)=XG(X)^2+u(X)^2,
	\]
	where $G(X)$ is monic of degree $r$ and $\deg u(X)\le r$. The polynomial $G(X)$ need not be squarefree.
	If $2^m\ge3r+3$, then there exists $\xi\in\mathbb F_{2^m}$ with $F(\xi^2)\ne0$ such that at least $2^m-3r-2$ values of $\eta\in\mathbb F_{2^m}$ satisfying $\eta\ne\xi$ and $F(\eta^2)\ne0$ give distinct monic squarefree polynomials
	\begin{equation}\label{eq:degree-extension}
		\widetilde G(X)=(X+\xi\eta)G(X)+(\xi+\eta)u(X)
	\end{equation}
	of degree $r+1$. For each such polynomial, let $L=\{\alpha\in\mathbb F_{2^m}:\widetilde G(\alpha)\ne0\}$. Then $d(\Gamma_2(L,\widetilde G))=2r+3$.
\end{theorem}

\begin{proof}
	Since $F(X)\mid X^{2^m}-X$, the polynomial $F(X)$ has $2r+1$ distinct roots in $\mathbb F_{2^m}$. Also, $F'(X)=G(X)^2$ and $F(X)$ is squarefree, so $\gcd(G(X),u(X))=1$.
	
	We first choose $\xi$ such that $F(\xi^2)\ne0$ and $\xi$ differs from the coefficient of $X^r$ in $u(X)$. Since $x\mapsto x^2$ is a bijection on $\mathbb F_{2^m}$, there are exactly $2^m-2r-1$ values of $\xi$ satisfying $F(\xi^2)\ne0$. The assumption $2^m\ge3r+3$ ensures that such a choice is possible.
	
	Let
	\[
	A(X)=XG(X)+\xi u(X),\qquad
	B(X)=\xi G(X)+u(X).
	\]
	Then $A(X)$ is monic of degree $r+1$, while $B(X)$ has degree $r$ by the choice of $\xi$. We show that $\gcd(A(X),B(X))=1$. Since $\gcd(G(X),u(X))=1$, we have $\gcd(G(X),B(X))=1$. Moreover,
	\[
	A(X)+\xi B(X)=(X+\xi^2)G(X).
	\]
	Thus every common root of $A(X)$ and $B(X)$ must equal $\xi^2$. But $B(\xi^2)^2=F(\xi^2)\ne0$, so $A(X)$ and $B(X)$ are relatively prime.
	
	We next determine the values of $\eta$ for which $\widetilde G(X)=A(X)+\eta B(X)$ is squarefree. Define
	\[
	D(X)=A'(X)B(X)+A(X)B'(X).
	\]
	The coefficient of $X^{2r}$ in $D(X)$ equals the leading coefficient of $B(X)$, which is nonzero. Hence $\deg D(X)=2r$. Since $D'(X)=0$, the polynomial $D(X)$ is a nonzero square and has at most $r$ distinct roots.
	
	Suppose that $\widetilde G(X)$ has a multiple root $\beta$ in an algebraic closure of $\mathbb F_{2^m}$. Since $\gcd(A(X),B(X))=1$, we have $B(\beta)\ne0$. From $\widetilde G(\beta)=\widetilde G'(\beta)=0$, we obtain
	\[
	D(\beta)=0,\qquad
	\eta=\frac{A(\beta)}{B(\beta)}.
	\]
	Thus each root of $D(X)$ determines at most one value of $\eta$ for which $\widetilde G(X)$ has a multiple root. Among the $2^m-2r-2$ values satisfying $\eta\ne\xi$ and $F(\eta^2)\ne0$, at most $r$ are unsuccessful. Therefore, at least $2^m-3r-2$ values give squarefree polynomials. Each $\widetilde G(X)$ is monic of degree $r+1$, and distinct values of $\eta$ give distinct polynomials since $B(X)\ne0$.
	
	For each such $\eta$, let
	\[
	\widetilde F(X)=(X+\xi^2)(X+\eta^2)F(X).
	\]
	The conditions on $\xi$ and $\eta$ imply that $\widetilde F(X)\mid X^{2^m}-X$ and $\deg\widetilde F(X)=2r+3$. Moreover,
	\[
	\begin{aligned}
		\widetilde F'(X)
		&=(\xi+\eta)^2F(X)+(X+\xi^2)(X+\eta^2)G(X)^2\\
		&=\bigl((X+\xi\eta)G(X)+(\xi+\eta)u(X)\bigr)^2\\
		&=\widetilde G(X)^2.
	\end{aligned}
	\]
	Since $\widetilde F(X)$ is squarefree, $\gcd(\widetilde F(X),\widetilde G(X))=1$. Hence all roots of $\widetilde F(X)$ belong to $L$, and Theorem~\ref{thm:split-locator} gives $d(\Gamma_2(L,\widetilde G))=2r+3$.
\end{proof}

\begin{remark}\label{rem:degree-extension-iteration}
	Since $\widetilde F(X)\mid X^{2^m}-X$ and $\deg\widetilde F(X)=2(r+1)+1$, the construction can be repeated whenever $2^m\ge3r+6$. This gives squarefree Goppa polynomials of every degree from $r+1$ to $\lfloor2^m/3\rfloor$ over $\mathbb F_{2^m}$, with nested root sets of their locator polynomials.
\end{remark}

\begin{corollary}\label{cor:uniform-existence}
	Let $m\ge1$ and $2\le r\le\lfloor2^m/3\rfloor$. There exist at least $2^m-3r+1$ distinct monic squarefree polynomials $G(X)\in\mathbb F_{2^m}[X]$ of degree $r$ such that, for $L=\{\alpha\in\mathbb F_{2^m}:G(\alpha)\ne0\}$, the binary Goppa code $\Gamma_2(L,G)$ has minimum distance $2r+1$.
\end{corollary}

\begin{proof}
	Choose $S\subseteq\mathbb F_{2^m}$ with $|S|=2r-1$, and let
	$
	F(X)=\prod_{\alpha\in S}(X-\alpha).
	$
	Then $F(X)$ is monic of degree $2r-1$ and satisfies $F(X)\mid X^{2^m}-X$. Since $2^m\ge3r$, Theorem~\ref{thm:degree-extension}, with $r-1$ in place of $r$, gives at least $2^m-3(r-1)-2=2^m-3r+1$ distinct monic squarefree Goppa polynomials of degree $r$, each defining a binary Goppa code with minimum distance $2r+1$.
\end{proof}

\begin{example}\label{ex:extension-F64}
	Let $\theta\in\mathbb F_{64}$ be a primitive element satisfying $\theta^6+\theta+1=0$. Let
	\[
	\begin{aligned}
		F(X)&=(X^7+1)(X^2+X+1),\\
		G(X)&=X^4+X^3+1,\qquad u(X)=X^4+X+1.
	\end{aligned}
	\]
	Then $F(X)=XG(X)^2+u(X)^2$ and $F(X)\mid X^{64}-X$. The code $\Gamma_2(\mathbb F_{64},G)$ has parameters $[64,40,9]_2$. Let $\xi=\theta$ and $\eta=\theta+1$. We have $\xi\ne\eta$ and $F(\xi^2)F(\eta^2)\ne0$. Equation~\eqref{eq:degree-extension} gives
	\[
	\begin{aligned}
		\widetilde G(X)
		&=X^5+(\theta^2+\theta)(X^4+X^3)+\theta^2+\theta+1,\\
		\widetilde F(X)
		&=(X+\theta^2)(X+\theta^2+1)F(X).
	\end{aligned}
	\]
	We have $\gcd(\widetilde G(X),\widetilde G'(X))=1$, and the only root of $\widetilde G(X)$ in $\mathbb F_{64}$ is $\theta^{46}$. Let $L=\mathbb F_{64}\setminus\{\theta^{46}\}$. The code $\Gamma_2(L,\widetilde G)$ has parameters $[63,33,11]_2$.
\end{example}

We next apply the construction to the polynomials
$P(X)=XW(X)^4+1$ in Proposition~\ref{prop:xw4-deletion}.

\begin{corollary}\label{cor:adjoining-xw4}
	Let $m\ge1$, and let $W(X)\in\mathbb F_{2^m}[X]$ be monic of degree $t\ge1$. Suppose that $P(X)=XW(X)^4+1$ is squarefree and splits over $\mathbb F_{2^m}$. For every $\eta\in\mathbb F_{2^m}^*$ satisfying $P(\eta^2)\ne0$, let
	\[
	G(X)=XW(X)^2+\eta
	\]
	and $L=\{\alpha\in\mathbb F_{2^m}:G(\alpha)\ne0\}$. Then $G(X)$ is monic and squarefree of degree $2t+1$, and $d(\Gamma_2(L,G))=4t+3$. This construction gives exactly $2^m-4t-2$ distinct Goppa polynomials.
\end{corollary}

\begin{proof}
	Since $P'(X)=W(X)^4$ and $\gcd(P(X),W(X))=1$, the polynomial $P(X)$ is squarefree. Let
	\[
	F(X)=X(X+\eta^2)P(X).
	\]
	Since $P(0)=1$, $\eta\ne0$, and $P(\eta^2)\ne0$, the polynomial $F(X)$ is monic of degree $4t+3$, is squarefree, and splits over $\mathbb F_{2^m}$. Hence $F(X)\mid X^{2^m}-X$. Moreover,
	\[
	\begin{aligned}
		F'(X)
		&=\eta^2P(X)+X(X+\eta^2)W(X)^4\\
		&=X^2W(X)^4+\eta^2\\
		&=G(X)^2.
	\end{aligned}
	\]
	Since $G'(X)=W(X)^2$ and $G(X)=\eta\ne0$ at every root of $W(X)$, we have $\gcd(G(X),G'(X))=1$. Thus $G(X)$ is squarefree. Also, $\gcd(F(X),G(X))=1$, so all roots of $F(X)$ belong to $L$. Theorem~\ref{thm:split-locator} gives $d(\Gamma_2(L,G))=4t+3$.
	
	Finally, $P(X)$ has $4t+1$ distinct roots in $\mathbb F_{2^m}$. Since $x\mapsto x^2$ is a bijection on $\mathbb F_{2^m}$, exactly $4t+1$ nonzero values of $\eta$ satisfy $P(\eta^2)=0$. Therefore, exactly $2^m-4t-2$ values of $\eta\in\mathbb F_{2^m}^*$ satisfy $P(\eta^2)\ne0$, and they give distinct Goppa polynomials.
\end{proof}

\section{Explicit Infinite Families of Binary Goppa Codes with Minimum Distance $2r+1$}
\label{sec:explicit-families}

In this section, we apply the constructions in Section~\ref{sec:constructions} to obtain explicit infinite families of binary Goppa codes with minimum distance $2r+1$. Throughout this section, let $L=\{\alpha\in\mathbb F_{2^m}:G(\alpha)\ne0\}$ for each Goppa polynomial $G(X)$.

\subsection{Families from Root Deletion}\label{subsec:root-deletion-families}

We first apply Theorem~\ref{thm:root-deletion} to subspace polynomials.

\begin{proposition}\label{prop:subspace-deletion}
	Let $2\le s\le m$, and let $V$ be an $s$-dimensional $\mathbb F_2$-subspace of $\mathbb F_{2^m}$. Write
	\[
	H(X)=\prod_{v\in V}(X-v)=\sum_{j=0}^{s}c_jX^{2^j}.
	\]
	If $c_1\ne0$, then
	\[
	G(X)=\sum_{j=1}^{s}c_j^{1/2}X^{2^{j-1}-1}
	\]
	is monic and squarefree of degree $2^{s-1}-1$, and $d(\Gamma_2(L,G))=2^s-1$.
\end{proposition}

\begin{proof}
	By Lemma~\ref{lem:subspace-polynomial}, the polynomial $H(X)$ has $2^s$ distinct roots in $\mathbb F_{2^m}$ and satisfies $H'(X)=c_0\ne0$. Let
	\[
	B(X)=\sum_{j=1}^{s}c_j^{1/2}X^{2^{j-1}}.
	\]
	Then $H(X)=c_0X+B(X)^2$ and $B'(X)=c_1^{1/2}\ne0$. Since $B'(X)$ is a nonzero constant, every root of $H(X)$ can be deleted by Theorem~\ref{thm:root-deletion}. Deleting the root $0$ yields $G(X)=B(X)/X$, which is monic and squarefree of degree $2^{s-1}-1$. Theorem~\ref{thm:root-deletion} also gives $d(\Gamma_2(L,G))=2^s-1$.
\end{proof}

Applying Proposition~\ref{prop:subspace-deletion} to absolute and relative trace polynomials gives the following two families.

\begin{proposition}\label{prop:absolute-trace-family}
	Let $m\ge1$ and $h\ge3$ with $h\mid m$. Then the polynomial $G(X)=\sum_{j=0}^{h-2}X^{2^j-1}$ is monic and squarefree of degree $2^{h-2}-1$, and $d(\Gamma_2(L,G))=2^{h-1}-1$.
\end{proposition}

\begin{proof}
	The polynomial $H(X)=\sum_{j=0}^{h-1}X^{2^j}$ is the trace polynomial of $\mathbb F_{2^h}$ over $\mathbb F_2$. Its roots form the $(h-1)$-dimensional $\mathbb F_2$-subspace $\{x\in\mathbb F_{2^h}:\operatorname{Tr}_{\mathbb F_{2^h}/\mathbb F_2}(x)=0\}$, which is contained in $\mathbb F_{2^m}$ since $h\mid m$. The coefficient of $X^2$ in $H(X)$ is $1$. The conclusion follows from Proposition~\ref{prop:subspace-deletion} with $s=h-1$.
\end{proof}

\begin{proposition}\label{prop:relative-trace-family}
	Let $m\ge1$ and $h\ge2$ with $2h\mid m$. For every $b\in\mathbb F_{2^h}^*$, the polynomial $G(X)=X^{2^h-1}+bX^{2^{h-1}-1}+1$ is monic and squarefree of degree $2^h-1$, and $d(\Gamma_2(L,G))=2^{h+1}-1$.
\end{proposition}

\begin{proof}
	The relative trace map $x\mapsto x^{2^h}+x$ from $\mathbb F_{2^{2h}}$ onto $\mathbb F_{2^h}$ has kernel $\mathbb F_{2^h}$. Since $b\ne0$, the polynomial
	\[
	H(X)=(X^{2^h}+X)(X^{2^h}+X+b^2)
	\]
	has $2^{h+1}$ distinct roots in $\mathbb F_{2^{2h}}$. Moreover,
	\[
	H(X)=X^{2^{h+1}}+b^2X^{2^h}+X^2+b^2X.
	\]
	Thus $H(X)$ is the subspace polynomial of an $(h+1)$-dimensional $\mathbb F_2$-subspace of $\mathbb F_{2^m}$. Its coefficient of $X^2$ is $1$, so Proposition~\ref{prop:subspace-deletion} applies with $s=h+1$.
\end{proof}

\begin{example}
	Let $m=8$, $h=4$, and let $\omega\in\mathbb F_4$ satisfy $\omega^2+\omega+1=0$. In Proposition~\ref{prop:relative-trace-family}, choose $b=\omega$. Deleting the root $0$ gives $G_0(X)=X^{15}+\omega X^7+1$. Since $H(X+1)=H(X)$, deleting the root $1$ gives
	\[
	G(X)=G_0(X+1)
	=\sum_{j=8}^{15}X^j+\omega^2\sum_{j=1}^{7}X^j+\omega.
	\]
	The binary Goppa code $\Gamma_2(L,G)$ has parameters $[256,136,31]_2$.
\end{example}

\begin{example}\label{ex:subspace-deletion}
	Table~\ref{tab:subspace-short-codes} gives short codes from Propositions~\ref{prop:subspace-deletion}--\ref{prop:relative-trace-family}. For $m=6$ and $m=8$, let $\mathbb F_{2^m}=\mathbb F_2(\theta)$, where $\theta$ satisfies $\theta^6+\theta+1=0$ and $\theta^8+\theta^4+\theta^3+\theta^2+1=0$, respectively. For the subspace constructions, choose $V=\langle1,\theta,\ldots,\theta^{s-1}\rangle_{\mathbb F_2}$. The absolute and relative trace constructions use $h=4$ and $(h,b)=(3,1)$, respectively.
\end{example}

\begin{table*}[t]
	\centering
	\caption{Short binary Goppa codes from subspace and trace polynomials}
	\label{tab:subspace-short-codes}
	\small
	\renewcommand{\arraystretch}{1.3}
	\setlength{\tabcolsep}{5pt}
	\begin{tabular}{|c|c|l|c|l|}
		\hline
		$m$ & $s$ & $G(X)$ & Parameters & Reference\\
		\hline
		$4$ & ---
		& $X^3+X+1$
		& $[16,4,7]_2$
		& Prop.~\ref{prop:absolute-trace-family}\\
		\hline
		
		$6$ & $3$
		& $X^3+\theta^{23}X+\theta^{17}$
		& $[64,46,7]_2$
		& Prop.~\ref{prop:subspace-deletion}\\
		\hline
		
		$6$ & $4$
		& $X^7+\theta^{13}X^3+\theta^{61}X+\theta^{31}$
		& $[63,21,15]_2$
		& Prop.~\ref{prop:subspace-deletion}\\
		\hline
		
		$8$ & $3$
		& $X^3+\theta^{15}X+\theta^{197}$
		& $[253,229,7]_2$
		& Prop.~\ref{prop:subspace-deletion}\\
		\hline
		
		$8$ & $4$
		& $X^7+\theta^{118}X^3+\theta^{44}X+\theta^{51}$
		& $[255,199,15]_2$
		& Prop.~\ref{prop:subspace-deletion}\\
		\hline
		
		$8$ & $5$
		& $X^{15}+\theta^{242}X^7+\theta^{182}X^3+\theta^{13}X+\theta^{212}$
		& $[255,135,31]_2$
		& Prop.~\ref{prop:subspace-deletion}\\
		\hline
		
		$6$ & ---
		& $X^7+X^3+1$
		& $[64,22,15]_2$
		& Prop.~\ref{prop:relative-trace-family}\\
		\hline
	\end{tabular}
\end{table*}

We next apply Proposition~\ref{prop:xw4-deletion} to three polynomials of the form $P(X)=XW(X)^4+1$. The splitting results for cases (ii) and (iii) follow from Zheng~\cite[proof of Theorem~9]{Zheng2026}.

\begin{proposition}\label{prop:xw4-families}
	Let $m\ge1$ and $h\ge2$. Consider the following polynomials $W(X)$ and conditions on $m$:
	\begin{enumerate}
		\item[(i)] $W(X)=X^{2^{h-2}}+1$, where $3h\mid m$;
		\item[(ii)] $W(X)=X^{2^{2h-2}}+X^{2^{2h-2}-2^{h-2}}$, where $5h\mid m$;
		\item[(iii)] $W(X)=X^{2^{3h-2}}+X^{2^{3h-2}-2^{h-2}}+X^{2^{3h-2}-2^{2h-2}}$, where $8h\mid m$.
	\end{enumerate}
	For every root $a\in\mathbb F_{2^m}$ of $XW(X)^4+1$, the polynomial
	\[
	G(X)=\frac{XW(X)^2+a}{X-a^2}
	\]
	is monic and squarefree of degree $2\deg W$, and $d(\Gamma_2(L,G))=4\deg W+1$. In each case, there are exactly $4\deg W+1$ distinct Goppa polynomials.
\end{proposition}

\begin{proof}
	Let $P(X)=XW(X)^4+1$. In case (i), we have $P(X)=X^{2^h+1}+X+1$. If $\beta$ is a root of $P(X)$, then $\beta\notin\{0,1\}$ and $\beta^{2^h}=(\beta+1)/\beta$. Raising this identity successively to the $2^h$-th power gives
	\[
	\beta^{2^{2h}}=\frac1{\beta+1},
	\qquad
	\beta^{2^{3h}}=\beta.
	\]
	Thus every root of $P(X)$ lies in $\mathbb F_{2^{3h}}$, so $P(X)$ splits over $\mathbb F_{2^{3h}}$.
	
	In cases (ii) and (iii), Zheng~\cite[proof of Theorem~9]{Zheng2026} proved that $P(X)$ splits over $\mathbb F_{2^{5h}}$ and $\mathbb F_{2^{8h}}$, respectively. Hence $P(X)$ splits over $\mathbb F_{2^m}$ in all three cases. Since $W(X)\in\mathbb F_2[X]$, every root $a$ of $P(X)$ satisfies $P(a^2)=P(a)^2=0$. Proposition~\ref{prop:xw4-deletion} therefore applies and gives the minimum distances and exact counts.
\end{proof}

\begin{example}\label{ex:xw4-families}
	Let $\omega\in\mathbb F_4$ satisfy $\omega^2+\omega+1=0$. We give one example for each family in Proposition~\ref{prop:xw4-families}.
	\begin{enumerate}
		\item[(i)] Let $h=2$, $m=6$, and $a=\omega$. Then $G(X)=X^2+\omega^2X+\omega^2$ has no roots in $\mathbb F_{64}$. The binary Goppa code $\Gamma_2(\mathbb F_{64},G)$ has parameters $[64,52,5]_2$.
		
		\item[(ii)] Let $h=2$, $m=10$, and $a=\omega$. Then
		$
		G(X)=X^8+\omega^2X^7+\omega^2X^6+\omega X^5+X^4+\omega^2X^3+\omega X^2+X+\omega^2.
		$
		The polynomial $G(X)$ has no roots in $\mathbb F_{1024}$. The binary Goppa code $\Gamma_2(\mathbb F_{1024},G)$ has parameters $[1024,944,17]_2$.
		
		\item[(iii)] Let $h=2$, $m=16$, and $a=1$. Then $G(X)=X^{32}+X^{31}+\sum_{j=0}^{24}X^j$ has exactly two roots $\omega$ and $\omega^2$ in $\mathbb F_{2^{16}}$. The binary Goppa code $\Gamma_2(L,G)$, where $L=\mathbb F_{2^{16}}\setminus\{\omega,\omega^2\}$, has parameters $[65534,65022,65]_2$.
	\end{enumerate}
\end{example}

\subsection{Families from Split Polynomials}\label{subsec:split-polynomial-families}
We first apply Theorem~\ref{thm:odd-polynomial} to a polynomial of degree five.

\begin{proposition}\label{prop:odd-mixed-family}
	Let $h\ge3$ and $2h\mid m$. Choose $z\in\mathbb F_{2^h}$ such that $b=z^2+z\notin\{0,1\}$. Then
	\[
	\begin{aligned}
		G(X)={}&X^{2^{h-1}+2}+(1+b)X^{2^{h-1}+1}+b^2X^{2^{h-1}}+\frac{b}{1+b}X^2+\frac{b^4}{1+b}
	\end{aligned}
	\]
	is monic and squarefree of degree $2^{h-1}+2$, and $d(\Gamma_2(L,G))=2^h+5$.
\end{proposition}

\begin{proof}
	Since $b=z^2+z\notin\{0,1\}$, the elements $0,1,z^2,z^2+1,b^2$ are distinct. Thus
	\[
	P(X)=X(X+1)(X+z^2)(X+z^2+1)(X+b^2)
	\]
	is monic and squarefree of degree five and splits over $\mathbb F_{2^h}$. Direct calculation gives $P(X)=A(X)^2+XB(X)^2$, where
	\[
	A(X)=bX^2+b^2X,\qquad
	B(X)=X^2+(1+b)X+b^2.
	\]
	Since $B'(X)=1+b\ne0$, the conditions of Theorem~\ref{thm:odd-polynomial} hold.
	
	Choose $c=b^2/(1+b)$. Then $A(X)+(X^{2^{h-1}}+c)B(X)$ equals the stated polynomial $G(X)$. Moreover,
	\[
	G'(X)=(1+b)X^{2^{h-1}},\qquad
	G(0)=\frac{b^4}{1+b}\ne0.
	\]
	Hence $\gcd(G(X),G'(X))=1$, so $G(X)$ is squarefree.
	
	Let $F(X)=(X^{2^h}+X+c^2)P(X)$. Since $c\ne0$, the polynomial $F(X)$ is squarefree and splits over $\mathbb F_{2^m}$. By the identity in the proof of Theorem~\ref{thm:odd-polynomial}, we have $F'(X)=G(X)^2$. Therefore, Theorem~\ref{thm:split-locator} gives $d(\Gamma_2(L,G))=2^h+5$.
\end{proof}

\begin{example}\label{ex:degree-interval}
	Table~\ref{tab:odd-degree-examples} gives examples of Corollary~\ref{cor:degree-interval} for $h=3$ and $m=6$. Let $\zeta\in\mathbb F_8$ satisfy $\zeta^3+\zeta+1=0$, where $\zeta^3+\zeta+1=0$. For Goppa degrees $5,6,7$, choose the root sets of $P(X)$ to be $\{0,1,\zeta^2\}$, $\{0,1,\zeta,\zeta^2,\zeta^6\}$, and $\mathbb F_8\setminus\{\zeta\}$, respectively, with $c=1,\zeta^3,1$.
\end{example}

\begin{table*}[t]
	\centering
	\caption{Binary Goppa codes from split polynomials with odd degree over $\mathbb F_{64}$}
	\label{tab:odd-degree-examples}
	\small
	\renewcommand{\arraystretch}{1.3}
	\begin{tabular}{|c|l|c|}
		\hline
		$\deg G(X)$ & $G(X)$ & Parameters\\
		\hline
		$5$ & $X^5+\zeta X^4+\zeta X+\zeta$ & $[63,33,11]_2$\\
		\hline
		$6$ & $X^6+(\zeta^2+\zeta+1)X^5+\zeta X^4+(\zeta^2+1)X^2+\zeta^2+\zeta$ & $[64,28,13]_2$\\
		\hline
		$7$ & $X^7+\zeta X^6+\zeta^2X^5+\zeta^3X^4+\zeta^5X^3+\zeta^6X^2+X+\zeta^3$ & $[64,22,15]_2$\\
		\hline
	\end{tabular}
\end{table*}
We next construct a family of four-term Goppa polynomials using a polynomial with six distinct roots over $\mathbb F_{16}$.

\begin{proposition}\label{prop:orbit-explicit}
	Let $m\ge1$, and let $v\ge1$ be odd with $15v\mid2^m-1$. Choose $\omega\in\mathbb F_4$ satisfying $\omega^2+\omega+1=0$. For each $c\in\{\omega,\omega^2\}$, the polynomial
	\[
	G(X)=X^{3v}+cX^{2v}+cX^v+1
	\]
	is monic and squarefree of degree $3v$, and $d(\Gamma_2(L,G))=6v+1$.
\end{proposition}

\begin{proof}
	Since $c^2+c+1=0$, the polynomial $R(X)=(X^2+X+1)^3+c^2X^2(X+1)^2$ factors as
	\begin{equation}\label{eq:four-term-factorization}
		R(X)=(X^2+X+c^2)(X^2+cX+c)(X^2+cX+1).
	\end{equation}
	Each quadratic factor splits over $\mathbb F_{16}$. Moreover,
	\[
	R'(X)=(X^2+X+1)^2,\qquad
	R(X)\equiv c^2\pmod{X^2+X+1}.
	\]
	Hence $R(X)$ is squarefree and has six distinct nonzero roots in $\mathbb F_{16}$.
	
	Since $15v\mid2^m-1$, every element of $\mathbb F_{16}^*$ has exactly $v$ distinct $v$-th roots in $\mathbb F_{2^m}$. Thus
	\begin{equation}\label{eq:four-term-locator}
		F(X)=XR(X^v)
	\end{equation}
	is monic and squarefree of degree $6v+1$ and splits over $\mathbb F_{2^m}$. As $v$ is odd,
	\[
	\begin{aligned}
		F'(X)
		&=R(X^v)+X^vR'(X^v)\\
		&=(X^{3v}+cX^{2v}+cX^v+1)^2\\
		&=G(X)^2.
	\end{aligned}
	\]
	
	We next show that $G(X)$ is squarefree. Since $G(0)=1$ and $G'(X)=X^{v-1}(X^{2v}+c)$, any common root $\beta$ of $G(X)$ and $G'(X)$ must satisfy $\beta\ne0$ and $\beta^{2v}=c$. But then $G(\beta)=c^2+1\ne0$, a contradiction. Therefore $\gcd(G(X),G'(X))=1$.
	
	Since $F(X)$ is squarefree and $F'(X)=G(X)^2$, we have $\gcd(F(X),G(X))=1$. All roots of $F(X)$ therefore belong to $L$, and Theorem~\ref{thm:split-locator} gives $d(\Gamma_2(L,G))=6v+1$.
\end{proof}

Proposition~\ref{prop:orbit-explicit} is also an instance of \cite[Theorem~3.2]{HeLiQu2026}, with $f(X)=X^3+cX^2+cX+1$, $t=v$, and the auxiliary polynomial $R(X)$ in \eqref{eq:four-term-factorization}. The factorization gives an explicit splitting condition $15v\mid2^m-1$.

\begin{example}\label{ex:orbit-count}
	Table~\ref{tab:four-term-examples} gives examples for
	Proposition~\ref{prop:orbit-explicit} with $c=\omega$, where $\omega^2+\omega+1=0$.
\end{example}

\begin{table*}[t]
	\centering
	\caption{Parameters of binary Goppa codes with
		$G(X)=X^{3v}+\omega X^{2v}+\omega X^v+1$}
	\label{tab:four-term-examples}
	\renewcommand{\arraystretch}{1.3}
	\small
	\begin{tabular}{|c|c|l|c|}
		\hline
		$m$ & $v$ & $G(X)$ & Parameters\\
		\hline
		$4$ & $1$ & $X^3+\omega X^2+\omega X+1$ & $[13,2,7]_2$\\
		\hline
		$8$ & $17$ & $X^{51}+\omega X^{34}+\omega X^{17}+1$ & $[205,2,103]_2$\\
		\hline
		$8$ & $1$ & $X^3+\omega X^2+\omega X+1$ & $[253,229,7]_2$\\
		\hline
	\end{tabular}
\end{table*}

\subsection{Families from Adjoining Roots}\label{subsec:adjoining-families}

Applying Corollary~\ref{cor:adjoining-xw4} to the polynomials in Proposition~\ref{prop:xw4-families} gives three further families with Goppa degrees increased by one.

\begin{proposition}\label{prop:two-root-families}
	Let $m\ge1$, $h\ge2$, and $\eta\in\mathbb F_{2^m}^*$. Consider the following polynomials:
	\begin{enumerate}
		\item[(i)] If $3h\mid m$, let
		\[
		G(X)=X^{2^{h-1}+1}+X+\eta.
		\]
		\item[(ii)] If $5h\mid m$, let
		\[
		G(X)=X^{2^{2h-1}+1}+X^{2^{2h-1}-2^{h-1}+1}+\eta.
		\]
		\item[(iii)] If $8h\mid m$, let
		\[
		G(X)=X^{2^{3h-1}+1}+X^{2^{3h-1}-2^{h-1}+1}+X^{2^{3h-1}-2^{2h-1}+1}+\eta.
		\]
	\end{enumerate}
	If $G(\eta^2)\ne0$, then $G(X)$ is monic and squarefree, and $d(\Gamma_2(L,G))=2r+1$, where $r=\deg G(X)$. In each case, exactly $2^m-2r$ values of $\eta$ satisfy this condition, giving distinct Goppa polynomials.
\end{proposition}

\begin{proof}
	In each case, let $W(X)$ be the corresponding polynomial in Proposition~\ref{prop:xw4-families}. Then $P(X)=XW(X)^4+1$ splits over $\mathbb F_{2^m}$ and $G(X)=XW(X)^2+\eta$. Moreover,
	\[
	G(\eta^2)^2=\eta^2P(\eta^2).
	\]
	Since $\eta\ne0$, the condition $G(\eta^2)\ne0$ is equivalent to $P(\eta^2)\ne0$. By Corollary~\ref{cor:adjoining-xw4}, the polynomial $G(X)$ is squarefree and $d(\Gamma_2(L,G))=2r+1$. The number of admissible values of $\eta$ is $2^m-4\deg W-2=2^m-2r$. Distinct values give distinct polynomials $G(X)$.
\end{proof}

\begin{example}\label{ex:two-root-families}
	We give three examples of Proposition~\ref{prop:two-root-families}.
	
	\begin{enumerate}
		\item[(i)] Let $h=3$, $m=9$, and $\eta=1$. Then $G(X)=X^5+X+1$, and the binary Goppa code $\Gamma_2(L,G)$ has parameters $[509,464,11]_2$.
		\item[(ii)] Let $h=2$, $m=10$, and $\eta=1$. Then $G(X)=X^9+X^7+1$, and $\Gamma_2(L,G)$ has parameters $[1019,929,19]_2$.
		\item[(iii)] Let $h=2$, $m=16$, and choose $\omega\in\mathbb F_4$ satisfying $\omega^2+\omega+1=0$. For $\eta=\omega$, the polynomial $G(X)=X^{33}+X^{31}+X^{25}+\omega$ has no roots in $\mathbb F_{2^{16}}$. Hence $\Gamma_2(\mathbb F_{2^{16}},G)$ has parameters $[65536,65008,67]_2$.
	\end{enumerate}
\end{example}

Finally, we give successive applications of Theorem~\ref{thm:degree-extension} over a fixed finite field.

\begin{example}\label{ex:extension-F32}
	Let $\theta\in\mathbb F_{32}$ satisfy $\theta^5+\theta^2+1=0$. Set $F_1(X)=X(X+1)(X+\theta)$ and $G_1(X)=X+\theta^{16}$, so that $F_1'(X)=G_1(X)^2$. Starting from $F_1(X)$, we apply Theorem~\ref{thm:degree-extension} four times, using the resulting locator polynomial at each step to construct the next. Table~\ref{tab:extension-F32} lists the additional roots, Goppa polynomials, and code parameters. At every step, the locator polynomial satisfies $F(X)\mid X^{32}-X$ and $F'(X)=G(X)^2$, with $G(X)$ squarefree.
\end{example}

\begin{table*}[t]
	\centering
	\caption{Binary Goppa codes over $\mathbb F_{32}$ from successive two-root extensions}
	\label{tab:extension-F32}
	\small
	\renewcommand{\arraystretch}{1.3}
	\begin{tabular}{|c|c|l|c|}
		\hline
		$\deg G(X)$ & New roots of $\widetilde F(X)$ & $G(X)$ & Parameters\\
		\hline
		$1$ & --- & $X+\theta^{16}$ & $[31,26,3]_2$\\
		\hline
		$2$ & $\theta^2,\theta^5$ & $X^2+\theta^{11}X+\theta^4$ & $[32,22,5]_2$\\
		\hline
		$3$ & $\theta^6,\theta^{27}$ & $X^3+\theta^{24}X^2+\theta^{17}X+\theta^5$ & $[31,16,7]_2$\\
		\hline
		$4$ & $\theta^4,\theta^{10}$ & $X^4+\theta^4X^3+\theta^{27}X^2+\theta^{10}X+\theta^{12}$ & $[30,10,9]_2$\\
		\hline
		$5$ & $\theta^7,\theta^{22}$ & $X^5+\theta^{25}X^4+\theta^{15}X^3+\theta^{14}X^2+\theta^8X+\theta^{11}$ & $[31,6,11]_2$\\
		\hline
	\end{tabular}
\end{table*}

\section{Goppa Codes from Split Auxiliary Polynomials}
\label{sec:split-auxiliary}

Dong et al.~\cite{DongChenChenXieLi2026} studied Goppa polynomials of the form
\begin{equation}\label{eq:auxiliary-representation}
	G(X)=U(X)H(X)+V(X)H'(X),
\end{equation}
where $U(X),V(X)\in\mathbb F_{q^m}[X]$ and $H(X)\in\mathbb F_q[X]$ is monic and irreducible of degree $\deg G(X)+1$, with all its roots in the support $L$. We show that the irreducibility assumption can be replaced by requiring $H(X)$ to have distinct roots in $L$. We then construct an explicit family of $q$-ary Goppa codes attaining the Goppa bound.

\begin{theorem}\label{thm:split-auxiliary-qary}
	Let $\Gamma_q(L,G)$ be a $q$-ary Goppa code with $\deg G(X)=r\ge1$. Suppose that $H(X)\in\mathbb F_{q^m}[X]$ is monic of degree $r+1$, has distinct roots $\alpha_{i_1},\ldots,\alpha_{i_{r+1}}\in L$, and satisfies \eqref{eq:auxiliary-representation} for some $U(X),V(X)\in\mathbb F_{q^m}[X]$. Then $\Gamma_q(L,G)$ contains a codeword with support $\{i_1,\ldots,i_{r+1}\}$ if and only if
	\begin{equation}\label{eq:split-auxiliary-ratio}
		\frac{V(\alpha_{i_j})}{V(\alpha_{i_{r+1}})}
		\in\mathbb F_q^*,\qquad 1\le j\le r.
	\end{equation}
	In this case, $d(\Gamma_q(L,G))=r+1$.
\end{theorem}

\begin{proof}
	Since $H(X)$ has distinct roots in $L$, we have $H'(\alpha_{i_j})\ne0$ and $G(\alpha_{i_j})\ne0$. Equation~\eqref{eq:auxiliary-representation} gives
	\[
	V(\alpha_{i_j})=\frac{G(\alpha_{i_j})}{H'(\alpha_{i_j})}\ne0,
	\qquad 1\le j\le r+1.
	\]
	The parity-check matrix restricted to these $r+1$ positions has rank $r$ over $\mathbb F_{q^m}$. By Lemma~\ref{lem:lagrange}, its unique null vector normalized by $c_{i_{r+1}}=1$ has coordinates
	\[
	c_{i_j}=\frac{V(\alpha_{i_j})}{V(\alpha_{i_{r+1}})},
	\qquad 1\le j\le r.
	\]
	Thus a $q$-ary codeword with the prescribed support exists precisely when \eqref{eq:split-auxiliary-ratio} holds. The Goppa bound $d(\Gamma_q(L,G))\ge r+1$ then gives equality.
\end{proof}

For a fixed monic squarefree polynomial $H(X)$ of degree $r+1$, the representation \eqref{eq:auxiliary-representation} exists for every polynomial $G(X)$ of degree at most $r$. Indeed, since $\gcd(H(X),H'(X))=1$, there is a unique polynomial $V(X)$ of degree at most $r$ satisfying $V(X)H'(X)\equiv G(X)\pmod{H(X)}$. The polynomial $U(X)=(G(X)-V(X)H'(X))/H(X)$ then satisfies \eqref{eq:auxiliary-representation}. Thus the condition in \eqref{eq:split-auxiliary-ratio} determines whether the roots of $H(X)$ support a codeword.

For binary Goppa codes with squarefree $G(X)$, we use $\Gamma_2(L,G)=\Gamma_2(L,G^2)$ and consider
\begin{equation}\label{eq:binary-auxiliary-representation}
	G(X)^2=U(X)H(X)+V(X)H'(X),
\end{equation}
where $U(X),V(X),H(X)\in\mathbb F_{2^m}[X]$.

\begin{corollary}\label{cor:split-auxiliary-binary}
	Let $\Gamma_2(L,G)$ be a binary Goppa code, where $G(X)$ is monic and squarefree of degree $r\ge1$ and satisfies \eqref{eq:binary-auxiliary-representation}. Suppose that $H(X)$ is monic of degree $2r+1$ with distinct roots $\alpha_{i_1},\ldots,\alpha_{i_{2r+1}}\in L$. Then $\Gamma_2(L,G)$ contains a codeword with support $\{i_1,\ldots,i_{2r+1}\}$ if and only if
	\begin{equation}\label{eq:binary-auxiliary-ratio}
		\frac{V(\alpha_{i_j})}{V(\alpha_{i_{2r+1}})}=1,
		\qquad 1\le j\le2r.
	\end{equation}
	Equivalently, $H'(X)=G(X)^2$. In this case, $d(\Gamma_2(L,G))=2r+1$.
\end{corollary}

\begin{proof}
	Applying Theorem~\ref{thm:split-auxiliary-qary} to $\Gamma_2(L,G^2)=\Gamma_2(L,G)$ gives \eqref{eq:binary-auxiliary-ratio}, since $\mathbb F_2^*=\{1\}$. By Proposition~\ref{prop:codeword-derivative}, the prescribed support defines a codeword if and only if $G(X)^2\mid H'(X)$. Since both polynomials are monic of degree $2r$, this is equivalent to $H'(X)=G(X)^2$.
\end{proof}

Corollary~\ref{cor:split-auxiliary-binary} recovers Theorem~\ref{thm:split-locator} in the auxiliary-polynomial form. When $H'(X)=G(X)^2$, the representation with $\deg V(X)<\deg H(X)$ has $V(X)=1$ and $U(X)=0$. For $q>2$, the values of $V(X)$ at the roots of $H(X)$ need only belong to a common coset of $\mathbb F_q^*$.

We next use the norm map to construct an explicit family satisfying \eqref{eq:split-auxiliary-ratio}.

\begin{proposition}\label{prop:dong-norm-family}
	Let $q\ge4$ be a prime power, $m\ge2$, and $r=\frac{q^m-1}{q-1}$. Choose $a\in\mathbb F_{q^m}^*$ with $\eta=a^r\ne1$, and let $\rho\in\mathbb F_q\setminus\{0,1\}$. Define
	\begin{equation}\label{eq:dong-norm-family}
		\begin{aligned}
			G(X)&=X^r+(\rho-1)aX^{r-1}-\rho,\\
			H(X)&=(X-a)(X^r-1),\\
			V(X)&=1-\rho+\frac{2\rho-1}{\eta-1}(X^r-1).
		\end{aligned}
	\end{equation}
	Then $G(X)$ is monic and squarefree of degree $r$, and $H(X)$ is monic and squarefree of degree $r+1$. Moreover, \eqref{eq:auxiliary-representation} holds for some $U(X)\in\mathbb F_{q^m}[X]$. For $L=\{\alpha\in\mathbb F_{q^m}:G(\alpha)\ne0\}$, we have $d(\Gamma_q(L,G))=r+1$ and $|L|\ge q^m-q+3$. This construction gives exactly $(q-2)^2r$ distinct Goppa polynomials.
\end{proposition}

\begin{proof}
	The map $a\mapsto a^r$ is the norm map from $\mathbb F_{q^m}^*$ onto $\mathbb F_q^*$, with exactly $r$ preimages of each nonzero element. Since $\eta\ne1$, the roots of $H(X)$ are $a$ and the $r$ elements $\beta\in\mathbb F_{q^m}^*$ satisfying $\beta^r=1$. Since $a^r\ne1$, the $r+1$ roots of $H(X)$ are distinct.
	
	We first show that $G(X)$ is squarefree. Since $r=1+q+\cdots+q^{m-1}$, differentiation gives $G'(X)=X^{r-1}$. As $G(0)=-\rho\ne0$, we have $\gcd(G(X),G'(X))=1$.
	
	Next, \eqref{eq:dong-norm-family} gives
	\[
	V(a)=\rho,\qquad V(\beta)=1-\rho
	\quad\text{for }\beta^r=1.
	\]
	Direct calculation also gives
	\[
	\frac{G(a)}{H'(a)}=\rho,\qquad
	\frac{G(\beta)}{H'(\beta)}=1-\rho
	\quad\text{for }\beta^r=1.
	\]
	Hence $G(X)-V(X)H'(X)$ vanishes at every root of $H(X)$. Since $H(X)$ is squarefree, it divides $G(X)-V(X)H'(X)$, so \eqref{eq:auxiliary-representation} holds with $U(X)=(G(X)-V(X)H'(X))/H(X)$. Both $\rho$ and $1-\rho$ belong to $\mathbb F_q^*$. Therefore, Theorem~\ref{thm:split-auxiliary-qary} gives $d(\Gamma_q(L,G))=r+1$.
	
	We next bound the number of roots of $G(X)$ in $\mathbb F_{q^m}$. Suppose that $G(\beta)=0$. Since $G(0)\ne0$, we have $\beta\ne0$, and
	\begin{equation}\label{eq:dong-norm-root}
		\beta(\beta^r-\rho)=(1-\rho)a\beta^r.
	\end{equation}
	Now $\beta^r\in\mathbb F_q^*$. Equation~\eqref{eq:dong-norm-root} implies $\beta^r\ne\rho$. If $\beta^r=1$, then $\beta=a$, contradicting $a^r\ne1$. Thus $\beta^r$ has at most $q-3$ possible values. For each such value, \eqref{eq:dong-norm-root} determines $\beta$ uniquely. Hence $G(X)$ has at most $q-3$ roots in $\mathbb F_{q^m}$, giving $|L|\ge q^m-q+3$.
	
	Finally, there are $(q-2)r$ choices of $a$ with $a^r\ne1$ and $q-2$ choices of $\rho$. The constant coefficient of $G(X)$ determines $\rho$, and its coefficient of $X^{r-1}$ then determines $a$. Thus all $(q-2)^2r$ Goppa polynomials are distinct.
\end{proof}

\begin{example}\label{ex:dong-norm-small}
	Table~\ref{tab:dong-norm-examples} gives examples of Proposition~\ref{prop:dong-norm-family} with $m=2$ and $a=\alpha$. For $q=4,5,7$, choose $\alpha$ satisfying $\alpha^4+\alpha+1=0$ over $\mathbb F_{16}$, $\alpha^2+\alpha+2=0$ over $\mathbb F_{25}$, and $\alpha^2+\alpha+3=0$ over $\mathbb F_{49}$, respectively. In each case, $L=\{\beta\in\mathbb F_{q^2}:G(\beta)\ne0\}$.
\end{example}

\begin{table}[t]
	\centering
	\caption{Examples from Proposition~\ref{prop:dong-norm-family}}
	\label{tab:dong-norm-examples}
	\footnotesize
	\setlength{\tabcolsep}{4pt}
	\renewcommand{\arraystretch}{1.3}
	\begin{tabular}{|c|c|l|c|}
		\hline
		$q$ & $\rho$ & $G(X)$ & Parameters\\
		\hline
		$4$ & $\alpha^5$ & $X^5+\alpha^{11}X^4+\alpha^5$ & $[16,6,6]_4$\\
		\hline
		$5$ & $2$ & $X^6+\alpha X^5+3$ & $[24,12,7]_5$\\
		\hline
		$5$ & $4$ & $X^6+3\alpha X^5+1$ & $[25,13,7]_5$\\
		\hline
		$7$ & $2$ & $X^8+\alpha X^7+5$ & $[49,33,9]_7$\\
		\hline
	\end{tabular}
\end{table}

\section{Binary Goppa Codes with Minimum Distance $2r+2$}\label{sec:even-distance}

In this section, we construct binary Goppa codes with minimum distance $2r+2$, where $r=\deg G(X)$. By Theorem~\ref{thm:even-min-distance}, it suffices to find an even-weight Goppa code and a monic polynomial $F(X)$ of degree $2r+2$ with distinct roots in its support such that $F'(X)=\lambda G(X)^2$ for some $\lambda\in\mathbb F_{2^m}^*$.

Let $G(X)\in\mathbb F_{2^m}[X]$ be monic and squarefree of degree $r\ge1$ with $G(0)=0$. For $\tau\in\mathbb F_{2^m}^*$, define
\begin{equation}\label{eq:quartic-even-general-support}
	L=\left\{\alpha\in\mathbb F_{2^m}^*:G(\alpha)\ne0,\ \operatorname{Tr}_{\mathbb F_{2^m}/\mathbb F_2}\left(\frac{\tau}{\alpha}\right)=1\right\}.
\end{equation}
The polynomial $u(X)=\tau G(X)/X$ has degree $r-1$ and satisfies $\operatorname{Tr}_{\mathbb F_{2^m}/\mathbb F_2}(u(\alpha)/G(\alpha))=1$ for all $\alpha\in L$. By Proposition~\ref{prop:even-weight}, every codeword of $\Gamma_2(L,G)$ has even weight. Hence $d(\Gamma_2(L,G))\ge2r+2$ whenever the code is nonzero. The same holds for any subset of $L$.

The following lemma provides the roots needed for our construction.

\begin{lemma}\label{lem:reciprocal-trace-polynomial}
	Let $h\ge1$ and $2h\mid m$. The polynomial $T(X)=X^{2^h}+X^{2^h-1}+1$ has $2^h$ distinct roots in $\mathbb F_{2^{2h}}$, each satisfying $\alpha^{-2^h}+\alpha^{-1}=1$.
	Let $\tau\in\mathbb F_{2^m}^*$ satisfy $\gamma=\operatorname{Tr}_{\mathbb F_{2^m}/\mathbb F_{2^{2h}}}(\tau)\in\mathbb F_{2^h}$. Then, for every $\alpha\in\mathbb F_{2^{2h}}^*$,
	\[
	\operatorname{Tr}_{\mathbb F_{2^m}/\mathbb F_2}\left(\frac{\tau}{\alpha}\right)
	=\operatorname{Tr}_{\mathbb F_{2^h}/\mathbb F_2}
	\bigl(\gamma(\alpha^{-2^h}+\alpha^{-1})\bigr).
	\]
\end{lemma}

\begin{proof}
	The relative trace map $y\mapsto y^{2^h}+y$ from $\mathbb F_{2^{2h}}$ onto $\mathbb F_{2^h}$ has kernel $\mathbb F_{2^h}$. Hence $y^{2^h}+y=1$ has $2^h$ distinct nonzero solutions. Substituting $y=\alpha^{-1}$ gives
	\[
	\alpha^{-2^h}+\alpha^{-1}=1,
	\]
	which is equivalent to $T(\alpha)=0$.

	For $\alpha\in\mathbb F_{2^{2h}}^*$, the transitivity of the trace map gives
	\begin{align*}
		\operatorname{Tr}_{\mathbb F_{2^m}/\mathbb F_2}
		\left(\frac{\tau}{\alpha}\right)
		&=\operatorname{Tr}_{\mathbb F_{2^{2h}}/\mathbb F_2}
		\left(\frac{\gamma}{\alpha}\right)\\
		&=\operatorname{Tr}_{\mathbb F_{2^h}/\mathbb F_2}
		\bigl(\gamma(\alpha^{-1}+\alpha^{-2^h})\bigr).
	\end{align*}
\end{proof}

We now construct a locator polynomial $F(X)=T(X)Q(X)$, where $Q(X)$ has four distinct roots satisfying $\alpha^{-2^h}+\alpha^{-1}=c\ne1$.

\begin{proposition}\label{prop:quartic-even-general}
	Let $h\ge3$ and $2h\mid m$, and put $r=2^{h-1}+1$. Choose $c,e\in\mathbb F_{2^h}$ such that $c\notin\{0,1\}$, $c^2+c+1\ne0$, $e^2+(c+1)e+1=0$, and $\operatorname{Tr}_{\mathbb F_{2^h}/\mathbb F_2}(e/c^2)=1$. Let $\theta\in\mathbb F_{2^h}$ satisfy $\theta^2(c^2+c+1)=1$, and define
	\begin{equation}\label{eq:quartic-even-general-G}
		G(X)=X^r+X^{r-1}+\theta X^{r-2}+(\theta+1)X.
	\end{equation}
	Choose $\tau\in\mathbb F_{2^m}^*$ such that $\gamma=\operatorname{Tr}_{\mathbb F_{2^m}/\mathbb F_{2^{2h}}}(\tau)\in\mathbb F_{2^h}$ and
	\[
	\operatorname{Tr}_{\mathbb F_{2^h}/\mathbb F_2}(\gamma)
	=\operatorname{Tr}_{\mathbb F_{2^h}/\mathbb F_2}(\gamma c)=1.
	\]
	Then $G(X)$ is monic and squarefree of degree $r$, and the support in \eqref{eq:quartic-even-general-support} gives $d(\Gamma_2(L,G))=2r+2=2^h+4$.
\end{proposition}

\begin{proof}
	We first show that $G(X)$ is squarefree. Since $c\notin\{0,1\}$ and $\theta^2(c^2+c+1)=1$, we have $\theta\ne1$. Direct calculation gives
	\[
	G(X)+XG'(X)=X^{r-1},\qquad G'(0)=\theta+1\ne0.
	\]
	Thus $G(X)$ and $G'(X)$ have no common root. Since $G(0)=0$, the support in \eqref{eq:quartic-even-general-support} ensures that every codeword has even weight.
	
	Next, let $F(X)=T(X)Q(X)$, where
	\[
	Q(X)=X^4+c(c+1)X^3+(c^2+c+1)X^2+1.
	\]
	The equation $e^2+(c+1)e+1=0$ gives
	\[
	Y^4Q(1/Y)
	=(Y^2+cY+e)(Y^2+cY+e+c+1).
	\]
	By assumption,
	\[
	\operatorname{Tr}_{\mathbb F_{2^h}/\mathbb F_2}
	\left(\frac{e}{c^2}\right)
	=\operatorname{Tr}_{\mathbb F_{2^h}/\mathbb F_2}
	\left(\frac{e+c+1}{c^2}\right)=1,
	\]
	where the second equality follows from $\operatorname{Tr}_{\mathbb F_{2^h}/\mathbb F_2}(1/c+1/c^2)=0$. By Lemma~\ref{lem:quadratic-splitting}, both quadratic factors are irreducible over $\mathbb F_{2^h}$ and split over $\mathbb F_{2^{2h}}$. Their roots satisfy $y^{2^h}+y=c$. Since their constant terms differ by $c+1\ne0$, the factors have no common root.
	
	Consequently, $Q(X)$ has four distinct roots $\alpha\in\mathbb F_{2^{2h}}$ satisfying $\alpha^{-2^h}+\alpha^{-1}=c$. By Lemma~\ref{lem:reciprocal-trace-polynomial} and $c\ne1$, the polynomials $T(X)$ and $Q(X)$ have no common root. Thus $F(X)$ has $2r+2$ distinct roots in $\mathbb F_{2^{2h}}$.
	
	Differentiating $F(X)$ gives
	\[
	\begin{aligned}
		F'(X)
		&=(c^2+c+1)X^{2^h+2}
		+(c^2+c+1)X^{2^h}
		+X^{2^h-2}+c(c+1)X^2\\
		&=(c^2+c+1)G(X)^2.
	\end{aligned}
	\]
	Since $F(X)$ is squarefree, we have $\gcd(F(X),G(X))=1$.
	
	We next verify that all roots of $F(X)$ belong to $L$. For each such root $\alpha$, we have $\alpha^{-2^h}+\alpha^{-1}\in\{1,c\}$. By Lemma~\ref{lem:reciprocal-trace-polynomial} and the conditions on $\gamma$,
	\[
	\operatorname{Tr}_{\mathbb F_{2^m}/\mathbb F_2}
	\left(\frac{\tau}{\alpha}\right)
	=\operatorname{Tr}_{\mathbb F_{2^h}/\mathbb F_2}
	\bigl(\gamma(\alpha^{-2^h}+\alpha^{-1})\bigr)=1.
	\]
	Hence all roots of $F(X)$ lie in $L$, and Theorem~\ref{thm:even-min-distance} gives $d(\Gamma_2(L,G))=2r+2$.
\end{proof}

Since $c\notin\mathbb F_2$, the two trace conditions on $\gamma$ are independent and have $2^{h-2}$ solutions in $\mathbb F_{2^h}$. Each solution has $2^{m-2h}$ preimages under $\operatorname{Tr}_{\mathbb F_{2^m}/\mathbb F_{2^{2h}}}$. Thus the required $\tau$ exists whenever $c$ and $e$ satisfy the assumptions of Proposition~\ref{prop:quartic-even-general}. We give two explicit families.

\begin{proposition}\label{cor:four-term-even}
	Let $h\ge3$ and $2h\mid m$, and put $r=2^{h-1}+1$. Choose $\theta$ and $\tau\in\mathbb F_{2^m}^*$ as follows:
	\begin{enumerate}
		\item[(i)] Let $h\equiv3\pmod6$, let $\theta\in\mathbb F_8$ satisfy $\theta^3+\theta+1=0$, and let $\operatorname{Tr}_{\mathbb F_{2^m}/\mathbb F_{2^{2h}}}(\tau)=\theta^6$.
		\item[(ii)] Let $h\equiv4\pmod8$, let $\theta\in\mathbb F_{16}$ satisfy $\theta^4+\theta^3+1=0$, and let $\operatorname{Tr}_{\mathbb F_{2^m}/\mathbb F_{2^{2h}}}(\tau)=\theta^3$.
	\end{enumerate}
	Then $G(X)$ in \eqref{eq:quartic-even-general-G} is monic and squarefree of degree $r$, and the support in \eqref{eq:quartic-even-general-support} gives $d(\Gamma_2(L,G))=2r+2=2^h+4$.
\end{proposition}

\begin{proof}
	In (i), let $c=\theta$, $e=\theta^2$, and $\gamma=\theta^6$. Since $\theta^3+\theta+1=0$, we have
	\[
	c^2+c+1=\theta^5,\qquad
	\theta^2(c^2+c+1)=1,
	\]
	together with $e^2+(c+1)e+1=0$ and $e/c^2=1$. Moreover, $\gamma c=1$ and
	\[
	\operatorname{Tr}_{\mathbb F_8/\mathbb F_2}(1)
	=\operatorname{Tr}_{\mathbb F_8/\mathbb F_2}(\theta^6)=1.
	\]
	Since $h/3$ is odd, the transitivity of the trace map shows that all three trace conditions in Proposition~\ref{prop:quartic-even-general} are satisfied.
	
	In (ii), let $c=\theta^3$, $e=\theta^7$, and $\gamma=\theta^3$. The equation $\theta^4+\theta^3+1=0$ gives
	\[
	c^2+c+1=\theta^{13},\qquad
	\theta^2(c^2+c+1)=1,
	\]
	and $e^2+(c+1)e+1=0$, $e/c^2=\theta$. Also, $\gamma c=\theta^6$, and
	\[
	\operatorname{Tr}_{\mathbb F_{16}/\mathbb F_2}(\theta)
	=\operatorname{Tr}_{\mathbb F_{16}/\mathbb F_2}(\theta^3)
	=\operatorname{Tr}_{\mathbb F_{16}/\mathbb F_2}(\theta^6)=1.
	\]
	Since $h/4$ is odd, the required trace conditions follow by transitivity. Proposition~\ref{prop:quartic-even-general} applies in both cases.
\end{proof}

\begin{remark}\label{rem:even-root-deletion}
	Further constructions are possible by deleting roots of $T(X)$. Let $Q(X)$ be a monic divisor of $T(X)$, and let $F(X)=T(X)/Q(X)$. If $\deg F(X)=2r+2$ and $F'(X)=\lambda G(X)^2$ for some $\lambda\in\mathbb F_{2^m}^*$ and a monic squarefree polynomial $G(X)$ of degree $r$ with $G(0)=0$, then the roots of $F(X)$ lie in the support \eqref{eq:quartic-even-general-support} whenever $\operatorname{Tr}_{\mathbb F_{2^h}/\mathbb F_2}(\gamma)=1$. Theorem~\ref{thm:even-min-distance} then gives $d(\Gamma_2(L,G))=2r+2$.
\end{remark}

\begin{example}\label{ex:even-distance-small}
	Table~\ref{tab:even-distance-examples} gives examples of Proposition~\ref{prop:quartic-even-general} with $h=m/2$. For $m=6$, let $\theta\in\mathbb F_8$ satisfy $\theta^3+\theta+1=0$ and choose $(c,e)=(\theta,\theta^2)$. For $m=8$, let $\theta\in\mathbb F_{16}$ satisfy $\theta^4+\theta^3+1=0$ and choose $(c,e)=(\theta^3,\theta^7)$. In each row, $L$ is defined by \eqref{eq:quartic-even-general-support} using the specified $\tau$. The first two rows correspond to Proposition~\ref{cor:four-term-even}.
\end{example}

\begin{table*}[t]
	\centering
	\caption{Parameters of binary Goppa codes with minimum distance $2r+2$}
	\label{tab:even-distance-examples}
	\renewcommand{\arraystretch}{1.3}
	\small
	\begin{tabular}{|c|c|l|c|}
		\hline
		$m$ & $\tau$ & $G(X)$ & Parameters\\
		\hline
		$6$ & $\theta^6$
		& $X^5+X^4+\theta X^3+(\theta+1)X$
		& $[32,6,12]_2$\\
		\hline
		$8$ & $\theta^3$
		& $X^9+X^8+\theta X^7+(\theta+1)X$
		& $[128,56,20]_2$\\
		\hline
		$8$ & $\theta$
		& $X^9+X^8+\theta X^7+(\theta+1)X$
		& $[128,58,20]_2$\\
		\hline
	\end{tabular}
\end{table*}

\section{Summary and Concluding Remarks}\label{s8}

In this paper, we investigated the minimum distances of binary Goppa codes defined by squarefree polynomials. Using codeword locator polynomials, we characterized codewords of weights $2r+1$ and $2r+2$. We developed three general constructions for binary Goppa codes attaining the designed distance $2r+1$ and obtained explicit infinite families. In particular, our constructions cover every Goppa degree between consecutive powers of two and yield a quantitative existence bound for every degree $2\le r\le\lfloor 2^m/3\rfloor$ over $\mathbb F_{2^m}$. We also extended a construction of $q$-ary Goppa codes using split auxiliary polynomials and obtained binary Goppa codes with minimum distance exactly $2r+2$.

The results provide general methods for constructing Goppa codes with determined minimum distances.
Further problems include determining the number of minimum-weight codewords and investigating the combinatorial structure of their supports, particularly their intersection properties. It is also of interest to determine how many permutation-inequivalent codes arise from the constructions presented here.


\begin{thebibliography}{99}
	
	\bibitem{BardetMoraTillich2024} M.~Bardet, R.~Mora, and J.-P.~Tillich, ``Polynomial time key-recovery attack on high rate random alternant codes,'' \emph{IEEE Trans. Inf. Theory}, vol.~70, no.~6, pp.~4492--4511, Jun.~2024.
	
	\bibitem{BezzateevShekhunova1995} S.~V.~Bezzateev and N.~A.~Shekhunova, ``Subclass of binary Goppa codes with minimal distance equal to the design distance,'' \emph{IEEE Trans. Inf. Theory}, vol.~41, no.~2, pp.~554--555, Mar.~1995.
	
	\bibitem{BezzateevShekhunova2008} S.~V.~Bezzateev and N.~A.~Shekhunova, ``Chain of separable binary Goppa codes and their minimal distance,'' \emph{IEEE Trans. Inf. Theory}, vol.~54, no.~12, pp.~5773--5778, Dec.~2008.
	
	\bibitem{BezzateevShekhunova2019} S.~V.~Bezzateev and N.~A.~Shekhunova, ``Totally decomposed cumulative Goppa codes with improved estimations,'' \emph{Des. Codes Cryptogr.}, vol.~87, no.~2--3, pp.~569--587, 2019.
	
	\bibitem{Byrne2023} I.~Byrne, N.~Dodson, R.~Lynch, E.~Pab\'on-Cancel, and F.~Pi\~nero-Gonz\'alez, ``Improving the minimum distance bound of trace Goppa codes,'' \emph{Des. Codes Cryptogr.}, vol.~91, no.~8, pp.~2649--2663, 2023.
	
	\bibitem{ChenChenDingLao2026b} Y.~Chen, H.~Chen, C.~Ding, and H.~Lao, ``On the minimum distances of some families of Goppa codes and BCH codes,'' 2026, arXiv:2604.25354.
	
	\bibitem{Delsarte1975} P.~Delsarte, ``On subfield subcodes of modified Reed--Solomon codes,'' \emph{IEEE Trans. Inf. Theory}, vol.~21, no.~5, pp.~575--576, Sep.~1975.
	
	\bibitem{DongChenChenXieLi2026} S.~Dong, H.~Chen, Y.~Chen, Z.~Xie, and C.~Li, ``A criterion to determine true minimum distances of Goppa codes,'' 2026, arXiv:2607.23139.
	
	\bibitem{Feng1983} G.-L.~Feng, ``Generalization of lower bound on the minimum distance and upper bound on the number of parity check digits for Goppa codes,'' \emph{Acta Electron. Sinica}, vol.~11, no.~2, pp.~66--72, 1983.
	
	\bibitem{Finiasz2003} M.~Finiasz, ``Words of minimal weight and weight distribution of binary Goppa codes,'' in \emph{Proc. IEEE Int. Symp. Inf. Theory (ISIT)}, Yokohama, Japan, 2003, p.~70.
	
	\bibitem{Finiasz2003extended} M.~Finiasz, ``Words of minimal weight and weight distribution in binary Goppa codes,'' extended manuscript accompanying the ISIT 2003 paper. [Online]. Available: \url{https://finiasz.net/research/2003/finiasz-isit03-extended.pdf}. Accessed: Oct.~13, 2026.
	
	\bibitem{Goppa1970} V.~D.~Goppa, ``A new class of linear correcting codes,'' \emph{Probl. Inf. Transm.}, vol.~6, no.~3, pp.~207--212, 1970.
	
	\bibitem{Goppa1971} V.~D.~Goppa, ``A rational representation of codes and $(L,g)$-codes,'' \emph{Probl. Inf. Transm.}, vol.~7, no.~3, pp.~223--229, 1971.
	
	\bibitem{HeLiQu2026} T.~He, K.~Li, and L.~Qu, ``Minimum distances of binary Goppa codes and constructions with prescribed alternating automorphism groups,'' 2026, arXiv:2608.02436.
	
	\bibitem{HeLiQu2027} T.~He, K.~Li, and L.~Qu, ``Some infinite families of binary Goppa codes with determined minimum distances,'' \emph{Discrete Math.}, vol.~350, no.~1, Art.~no.~115405, 2027.
	
	\bibitem{LevyDitVehelLitsyn1997} F.~Levy-dit-Vehel and S.~Litsyn, ``Parameters of Goppa codes revisited,'' \emph{IEEE Trans. Inf. Theory}, vol.~43, no.~6, pp.~1811--1819, Nov.~1997.
	
	\bibitem{MacWilliamsSloane1977} F.~J.~MacWilliams and N.~J.~A.~Sloane, \emph{The Theory of Error-Correcting Codes}. Amsterdam, The Netherlands: North-Holland, 1977.
	
	\bibitem{MorenoMoreno1992} C.~J.~Moreno and O.~Moreno, ``Exponential sums and Goppa codes: II,'' \emph{IEEE Trans. Inf. Theory}, vol.~38, no.~4, pp.~1222--1229, Jul.~1992.
	
	\bibitem{Patterson1975} N.~J.~Patterson, ``The algebraic decoding of Goppa codes,'' \emph{IEEE Trans. Inf. Theory}, vol.~21, no.~2, pp.~203--207, Mar.~1975.
	
	\bibitem{SugiyamaKasaharaHirasawaNamekawa1976} Y.~Sugiyama, M.~Kasahara, S.~Hirasawa, and T.~Namekawa, ``Further results on Goppa codes and their applications to constructing efficient binary codes,'' \emph{IEEE Trans. Inf. Theory}, vol.~22, no.~5, pp.~518--526, Sep.~1976.
	
	\bibitem{Veron1998} P.~V\'eron, ``Goppa codes and trace operator,'' \emph{IEEE Trans. Inf. Theory}, vol.~44, no.~1, pp.~290--294, Jan.~1998.
	
	\bibitem{YueJi1991} D.-W.~Yue and Q.-J.~Ji, ``On the lower bounds on minimum distance for Goppa codes and for alternant codes,'' \emph{J. China Inst. Commun.}, vol.~12, no.~3, pp.~10--16, 1991.
	
	\bibitem{Zheng2026} R.~Zheng, ``Minimum distances of primitive narrow-sense BCH codes via good zero-sets,'' Sep.~2026, arXiv:2609.21994v2.
\end{thebibliography}
\end{document}